\documentclass[aps,pra,twocolumn,a4paper]{revtex4-1}
\usepackage{mathrsfs} 
\usepackage{mathrsfs,amsmath, amsthm, amssymb}
\usepackage{graphicx}
\usepackage{float}
\usepackage{booktabs}
\usepackage{multirow}
\usepackage{xcolor}
\usepackage{hyperref}
\hypersetup{colorlinks=true, citecolor=blue, urlcolor=blue, linkcolor=blue}
\begin{document}

\bibliographystyle{apsrev4-2}
\title{Fourth-order moment of the light field in atmosphere for moderate and strong turbulence}


%
%

\author{Roman Baskov}
\email{Corresponding address: oleksa$_$baskov@ukr.net}
\affiliation{Bogolyubov Institute for Theoretical Physics of the National Academy of Sciences of Ukraine,\\
	Metrolohichna Street 14-b, Kyiv 03143, Ukraine}
\affiliation{ Institute of Physics of the National Academy of Sciences of Ukraine,\\ prospekt Nauky 46, Kyiv 03028, Ukraine}


\begin{abstract}
Collisionless Boltzmann equation is used to describe the intensity correlations in partially saturated and fully saturated regimes in terms of photon distribution function in the phase space. Explicit expression for fourth moment of the light fields is obtained for the case of moderate and strong turbulence. Such expression consists of two terms accounting for two regions in the phase space that independently contribute to the correlation function.  It is shown that present solution agrees with previous results for fully saturated regime. Additionally it embodies the effect of partially saturated radiation where the correlations of photon trajectories are important and the magnitude of the scintillation index is well above the unity. Fourth moment is used to study the fluctuations of transmittance which consider the effect of finite detector aperture.
\end{abstract}

\maketitle

\section{Introduction}

Propagation of the light in atmospheric channel is an essential part of many new-age applied areas such as quantum key distribution \cite{villoresi,usenko}, satellite-ground communication \cite{Hosseinidehaj, Aspelmeyer}, quantum teleportation \cite{ma,ren,Hofmann}, propagation of entangled and squeezed states \cite{ursin,yin,Peuntinger,vasylev2016}. Altogether these areas contribute to the the development of the next-generation quantum and classical communication systems including quantum internet \cite{Meyers}, quantum protocols \cite{cao2020, kish}, etc. However, spatio-temporal properties of the light in atmosphere are modified drastically in a way of propagation limiting current applications. 

Fluctuations of refractive index in Earth's atmosphere introduce random distortions to the phase of waves. Since range of sizes of optical inhomogeneities is very wide, from millimeters (inner scale of turbulence, $l_0$) to hundreds of meters (outer scale of turbulence, $L_0$), laser beam is exposed to the bunch of various negative effects: beam spreading, beam wandering, fragmentation, beam jitter, intensity fluctuations, etc \cite{Fante1, Fante2, Tatarskii1, Andrews}. All of them affect statistical and spatio-temporal properties of the light radiation causing additional to absorption and scattering losses in atmosphere and impairing the performance of free-space communication systems. 

Although intensity fluctuations plays critical role in the transmission of optical signal, their description remains one of the most challenging problem in free-space optics. For atmospheric channel the initially coherent laser radiation acquires some properties of Gaussian statistics \cite{DeWolf}. This leads to saturation effect of the optical wave \cite{gracheva}. Such gradual change of statistical properties complicates theoretical analysis of correlation properties (fourth-order moment) of propagating radiation. 

Some general approach involving equations of evolution for fourth-order moments was proposed in early studies \cite{Fante1,Fante2}. However, the applicability of these equations is limited due to their complexity. Besides that correlations of intensity were studied in terms of scintillation index and covariance function \cite{Andrews, andrews1997, Dashen, banakh79}. Nevertheless, there is still no rigorous theory of fourth moment for optical fields in atmosphere for moderate and strong turbulence (partially saturated and fully saturated regimes).

Here we use method of photon distribution function (PDF) in the phase space \cite{berman2006} to describe laser radiation in atmosphere. PDF is defined as photon density in coordinate-momentum space (phase space). Intensity of the light and correlation properties of radiation are derived from PDF moments. Method of PDF was successfully applied to the problem of light propagation in lossy channels \cite{berman2009, Gorshkov, stolyarov2013,stolyarov2014}. Particularly it proved to be effective for study of intensity fluctuations (scintillation) in the range of moderate turbulence regime \cite{baskov2018, baskov2016} and fourth order moment problem in asymptotic case of large propagation distances, $z\rightarrow\infty$  \cite{baskov2020}.  In current paper, for the description of partially saturated and saturated regimes where only substantial change of photon momenta due to atmospheric turbulence takes place we use Boltzmann kinetic equation for PDF \cite{berman2006}. In this case the effect of turbulence is enclosed in random force originating from the gradient of refractive index. 
	
The paper is devoted to derivation of the expression for fourth moment applicable for the case of moderate-to-strong and strong turbulence regimes. Such function plays leading role in many applied researches including intensity correlation \cite{Vellekoop2010,Newman}, enhanced focusing \cite{Popoff,Vellekoop2007}, different imaging problems \cite{Katz, Hardy, Zhang_imag, wang, Shi}, reconstruction of the probability distribution of transmittance \cite{Semenov2009} and its applications to communication protocols \cite{Xue2020}. In paper we utilize the effect of multiple collisions with turbulent inhomogeneities which leads to the Gaussian statistics of radiation. In this case only particular volume in the phase space contribute to the correlation function.

The remainder of this paper is organized as follows. In Sec. \ref{sec:pdf}, we provide review of  photon distribution function approach applied to laser beam propagation in atmosphere. In Sec. \ref{sec:gamma4}, explicit expression for fourth moment of the light fields and its analysis are presented. Section \ref{sec:trans_fluct} devoted to estimation of transmittance fluctuations and its dependence on the size of detector aperture.  In Appendix \ref{appendix1}, we give detailed derivation of the expressions for fourth moment terms.

\section{Preliminaries}
\label{sec:pdf}

\subsection*{Photon distribution function.}

The photon distribution function resembling the idea of distribution functions 
in physics of solids \cite{Tom} is given by \cite{berman2006,ujp1992,*ujp2012}
\begin{equation}\label{1threee}
\hat{f}({\bf r},{\bf q},t)=\frac 1V\sum_{\bf k}e^{-i{\bf k\cdot r}}b^\dag_{{\bf
		q}+ {\bf k}/2}b_{{\bf q}-{\bf k}/2},
\end{equation}
where $b^\dag_{\bf q}$ and $b_{\bf q}$ are the quantum amplitudes of bosonic photon field with the wave vector ${\bf q}$; $V\equiv L_xL_yL_z\equiv SL_z$ is the normalizing volume.  All operators are given in the Heisenberg representation. The laser beam  propagates in the $z$ direction. It is assumed that ${\bf k}_\perp\,,{\bf q}_\perp\ll q_0$ where $q_0$ is the wave vector corresponding to the
central frequency $\omega_0$ of radiation $\omega_0=cq_0$, $c$ is the speed of light in a vacuum. Such assumption justifies the paraxial approximation.  The initial polarization of light left out of consideration in this case as for a wide range of propagation distances it remains almost constant (see Ref. \cite{stroh}).

The Hamiltonian of photons in a medium with a fluctuating
refractive index could be derived from representation of energy in inhomogeneous media \cite{landau}
\begin{equation}\label{one}
H=\sum_{\bf k}\hbar \omega _{\bf k}b^\dag_{\bf k}b_{\bf k}-
\sum_{{\bf k},{\bf k^\prime}}\hbar \omega _{\bf k}n_{\bf k^\prime}b^\dag_{\bf k}b_{{\bf k}
	+{\bf k^\prime}}
\end{equation} 
where $\hbar \omega _{\bf k}\equiv \hbar ck$ is the photon energy,
and $n_{\bf k}$ is the Fourier transform of the refractive index
fluctuations $\delta n({\bf r})$. The Fourier transform is defined by
\begin{equation}\label{two}
n_{\bf k}=\frac 1V\int dVe^{i{\bf kr}}\delta n({\bf r}).
\end{equation}
Usually, $\delta n$ is assumed to be a Gaussian random variable with known
covariance $\langle\delta n({\bf r})\delta n({\bf r}^\prime )\rangle$. The covariance is defined
by its Fourier transform, $\psi ({\bf g})$, with respect
to the difference ${\bf r}-{\bf r}^\prime$.
In a statistically homogeneous atmosphere it can be written as
\begin{equation}\label{ad}
\langle\delta n({\bf r}-{\bf r}^\prime)\delta n(0)\rangle=\int d{\bf g}
e^{-i{\bf g}(\bf {r-r^\prime})}\psi({\bf g}).
\end{equation}

The evolution equation for PDF is derived from Heisenberg's equation. Although in more general case its evolution is gathered by Boltzmann-Langevin equation \cite{baskov2018} which takes to account whole range of possible changes in photon momentum due to collisions with atmosphere inhomogeneities, for reasonably long distances (see \cite{baskov2016}) description with collisionless Boltzmann equation
\begin{equation}\label{6seven1}
\partial_t \hat{f}({\bf r},{\bf q},t)+{\bf c_q}\cdot\partial_{\bf r}\hat{f}({\bf r},{\bf q},t)+
{\bf F}({\bf r})\cdot\partial_{\bf q}\hat{f}({\bf r},{\bf q},t)=0
\end{equation}
is justified. In this case the effect of atmospheric turbulence is enclosed in random smooth force ${\bf F}({\bf r})=\omega_0\partial_{\bf r}n({\bf r})$. 

The general solution of (\ref{6seven1}) is obtained by characteristics method
\begin{equation}\label{nine}
\hat{f}({\bf r},{\bf q},t)=\phi \Bigg\{{\bf r}-\int _0^tdt^\prime\frac
{\partial {\bf r}(t^\prime)}{\partial t^\prime};{\bf
	q}-\int_0^tdt^\prime\frac {\partial {\bf q}(t^{\prime})}{\partial
	t^\prime}\Bigg\},
\end{equation}
where evolution of PDF is described in terms of the classical trajectories of photons 
\begin{eqnarray}\label{eight}
\frac {\partial {\bf r}(t^\prime)}{\partial t^\prime}={\bf c}[{\bf q}(t^\prime)],\\
\frac {\partial {\bf q}(t^\prime)}{\partial t^\prime}={\bf F}[{\bf
	r}(t^\prime)],
\end{eqnarray}
the function $\phi ({\bf r},{\bf q})$ is the ``initial" value
of $\hat{f}({\bf r},{\bf q},t)$ in the aperture plane of the source, i.e.
\begin{equation}\label{ten}
\phi ({\bf r},{\bf q})=\frac 1V\sum_{\bf k}e^{-i{\bf kr}}(b^+_{{\bf q}+
	\frac{\bf k}2}b_{{\bf q}-\frac{\bf k}2})|_{t=0}\equiv \sum_{\bf k}e^{-i{\bf kr}}
\phi ({\bf k},{\bf q}).
\end{equation}
Under paraxial approximation atmosphere mostly affects divergence of the beam, ${\bf q}_\perp$, and has negligible influence on the longitudinal components ($z$-axis).  
Then, Eq. (\ref{nine}) can be written as
\begin{eqnarray}\label{thirteen}
\hat{f}({\bf r},{\bf q},t){=}\phi \Bigg\{{\bf r}{-}{\bf c_q}t{+}\frac c{q_0}\int\limits _0^tdt^\prime
t^\prime {\bf F}_\bot [{\bf r}(t^\prime )];\hspace{10mm}\\
{\bf q}{-}\int\limits _0^tdt^\prime
{\bf F}_\bot [{\bf r}(t^\prime )]\Bigg\}.\nonumber
\end{eqnarray}
Assuming the initial configuration for laser radiation is known, the first and second moments of $\hat{f}$, which describe 
beam intensity and its correlations, could be obtained by means of the iterative procedure for powers of ${\bf F}_\perp({\bf r})$ \cite{berman2006,baskov2016}.

\section{Intensity correlations}
\label{sec:gamma4}

The intensity (density of photons in the spatial domain) is derived from operator $\hat{f}({\bf r},{\bf q},t)$ by summation over all values of $ \bf q $ 
\begin{equation}\label{2my}
\hat{I}({\bf r},t)=\sum_{\bf q}\hat{f}({\bf r},{\bf q},t)=\frac 1V\sum_{\bf q,k}e^{-i{\bf k\cdot r}}b^\dag_{{\bf
		q}+ {\bf k}/2}b_{{\bf q}-{\bf k}/2}.
\end{equation}
Consequently, the intensity correlations, fourth-order moment for the field operators, is defined by
\begin{eqnarray}\label{4my}
\Gamma_4({\bf r},{\bf r}')&\equiv&\langle\hat{I}({\bf r}, t)\hat{I}({\bf r'}, t)\rangle\\
&{=}&\frac{1}{V^2}\sum_{{\bf q,k},\atop{\bf q',k'}}e^{-i({\bf k}\cdot{\bf r}{+}{\bf k'}\cdot{\bf r'})}\langle b^\dag_{{\bf q}{+}\frac{\bf k}{2}}b_{{\bf q}{-}\frac{\bf k}{2}}b^\dag_{{\bf q'}{+}\frac{\bf k'}{2}}b_{{\bf q'}{-}\frac{\bf k'}{2}}\rangle.\nonumber
\end{eqnarray} 
\sloppy The averaging $\langle\dots\rangle$ includes the quantum-mechanical averaging of operators $\hat{I}$ and averaging over different configurations of atmospheric turbulence. Both averaging can be performed independently.

It was shown in the recent research \cite{baskov2020} that due to saturation effect for fluctuations in asymptotic case of large distances, $z\rightarrow \infty$,  fourth moment may be expressed via second moments
\begin{eqnarray}\label{5my}
\langle b^\dag_{{\bf q}{+}\frac{\bf k}{2}}b_{{\bf q}{-}\frac{\bf k}{2}}b^\dag_{{\bf q'}{+}\frac{\bf k'}{2}}b_{{\bf q'}{-}\frac{\bf k'}{2}}\rangle{\approx}\hspace{40mm}\\
\langle b _{{\bf q}{+}\frac{\bf k}{2}}^\dag b _{{\bf q}{-}\frac{\bf k}{2}}\rangle\langle b _{{\bf q}^\prime{+}\frac{{\bf k}^\prime}{2}}^\dag
b _{{\bf q}^\prime{-}\frac{{\bf k}^\prime}{2}}\rangle
{+}\langle b _{{\bf q}+\frac{\bf k}{2}}^\dag b _{{\bf q}^\prime -\frac{{\bf k}^\prime}{2}}\rangle\langle b _{{\bf q}^\prime+\frac{{\bf k}^\prime}{2}}^\dag
b _{{\bf q}{-}\frac{\bf k}{2}}\rangle\nonumber\\
=n_{\bf q}n_{{\bf q}^\prime}\delta _{{\bf k},0}\delta _{{\bf k}^{\prime },0} +n_{{\bf q}+\frac{\bf k}2}n_{{\bf q}-\frac{\bf k}2}\delta _{{\bf q},{\bf q}^\prime }\delta _{{\bf k},-{\bf k}^\prime },\nonumber
\end{eqnarray} 
where $n_{\bf q}\equiv \langle b^\dag_{\bf q}b_{\bf q}\rangle$. Also it is assumed that initial laser radiation is in a multiphoton coherent state, so the shot-noise term is omitted. Expression (\ref{5my}) is  legitimate if the amplitudes $b^\dag$ and $b$ obey Gaussian  statistics. In other words for $t\rightarrow\infty$ we assume that each primary coherent electromagnetic wave experiences multiple scatterings by randomly distributed turbulent eddies \cite{Dashen,DeWolf} and the radiation becomes fully saturated.

At large but finite $z$, in partially saturated regime, one should consider ``nondiagonal'' terms in four-wave correlations (\ref{5my}). In this case there are two regions where pair correlations of the field operators should be taken into account \cite{berman2006}: $(i)$ $k, k'\le R_b^{-1}$, $(ii)$ $|{\bf q}-{\bf q}'+({\bf k}+{\bf k}')/2|,|{\bf q}-{\bf q}'+({\bf k}+{\bf k}')/2|\le R_b^{-1}$, where $R_b^2\equiv \langle{\bf r}^2\rangle_T $ (see \cite{baskov2020}) is the ''turbulent" part of beam radius,  $R_b^2={8z^3c\alpha}/{(3r_0^2\omega_0^2)}$, $\alpha =0.5{\pi\omega_0^2}{c}^{-1}\int d{\bf g} g^{2}\psi(g)$. ``Turbulent'' term should be dominant to initial radius of beam and diffraction term, $R_b^2>r_0^2, 4z^2q_0^{-2}r_0^{-2}$, to distinguish curtain level of saturation of the fluctuations, i.e. partially saturated and fully saturated regime. Exploiting the approach from \cite{berman2006, baskov2016} fourth moments for these two regions are obtained for Gaussian beams (see Appendix \ref{appendix1} for details), $\Gamma_4({\bf r},{\bf r}')=\Gamma_4^{(i)}({\bf r},{\bf r}')+\Gamma_4^{(ii)}({\bf r},{\bf r}')$:
\begin{widetext} 
\begin{eqnarray}\label{gamma4_i}
\Gamma_4^{(i)}({\bf r},{\bf r}')=2\pi C\int\limits_{0}^\infty d\tilde{q}\tilde{q}\left[F_1^{\tilde{q},\rho_{\parallel}}F_2^{\tilde{q},\rho_{\parallel}}\big(F_3^{\tilde{q},\rho_{\parallel}}H_1^{\tilde{q},\rho_{\parallel}}-{G_1^{\tilde{q},\rho_{\parallel}}}^2\big)\big(F_4^{\tilde{q},\rho_{\parallel}}H_2^{\tilde{q},\rho_{\parallel}}-{G_2^{\tilde{q},\rho_{\parallel}}}^2\big)\right]^{-\frac{1}{2}}\times\hspace{30mm}\\\exp\left\{-\frac{\left(\tilde{q}-\rho_{\parallel}\frac{G_1^{\tilde{q},\rho_{\parallel}}}{2F_3^{\tilde{q},\rho_{\parallel}}}\right)^2}{(H_1^{\tilde{q},\rho_{\parallel}}-{G_1^{\tilde{q},\rho_{\parallel}}}^2/F_3^{\tilde{q},\rho_{\parallel}})}\right\}\exp\left\{-\frac{\rho_\perp^2H_2^{\tilde{q},\rho_{\parallel}}}{(H_2^{\tilde{q},\rho_{\parallel}}F_4^{\tilde{q},\rho_{\parallel}}-{G_2^{\tilde{q},\rho_{\parallel}}}^2)}\right\}
\exp\left\{-\left(\frac{\rho_{\parallel}^{\prime 2}}{F_1^{\tilde{q},\rho_{\parallel}}}+\frac{\rho_{\perp}^{\prime 2}}{F_2^{\tilde{q},\rho_{\parallel}}}+\frac{\rho_{\parallel}^{2}}{4F_3^{\tilde{q},\rho_{\parallel}}}\right)\right\}\nonumber,
\end{eqnarray}
\begin{eqnarray}\label{gamma4_ii}
\Gamma_4^{(ii)}({\bf r},{\bf r}')=2\pi C\int\limits_{0}^\infty d\tilde{q}\tilde{q}\left[F_1^{\tilde{q}}F_2^{\tilde{q}}\big(F_3^{\tilde{q}}H_1^{\tilde{q}}-{G_1^{\tilde{q}}}^{\,2}\big)\big(F_4^{\tilde{q}}H_2^{\tilde{q}}-{G_2^{\tilde{q}}}^{\,2}\big)\right]^{-\frac{1}{2}}\times\hspace{65mm}\\\exp\left\{-\frac{{\tilde{q}}^2}{H_1^{\tilde{q}}-{G_1^{\tilde{q}}}^{\,2}/F_3^{\tilde{q}}}\right\}
\exp\left\{-\left(\frac{\rho_{\parallel}^{\prime 2}}{F_1^{\tilde{q}}}+\frac{\rho_{\perp}^{\prime 2}}{F_2^{\tilde{q}}}\right)\right\}\exp\left\{-i2{\tilde{q}}\rho_{\parallel}\frac{q_0}{z}\right\}\nonumber,
\end{eqnarray}
\end{widetext}
where $\boldsymbol{\rho}={\bf r}-{\bf r}^\prime$, $\boldsymbol{\rho}^\prime=({\bf r}+{\bf r}^\prime)/2$ are two-dimensional vectors transverse to propagation direction $z$; $F$, $G$, $H$ are functions of $\rho_{\parallel}$ and ${\tilde{q}}$ for $(i)$ and functions of ${\tilde{q}}$ for $(ii)$ contributions to $\Gamma_4$; $C$ is constant derived from total flux. It is worth to emphasize that although intensity correlations are evaluated for two points ${\bf r}$ and ${\bf r}^\prime$, the expression for fourth moment is expressed via the difference ${\bf r}-{\bf r}^\prime$ and the center of mass position $({\bf r}+{\bf r}^\prime)/2$. The dependence on $\boldsymbol{\rho}^\prime$ has a simple Gaussian form. At the same time dependence on $\boldsymbol{\rho}$ is quite intricate and accounts for the correlations of different trajectories. Such a situation could be favorable for the calculation of integral quantities that consider spatial distribution of the radiation in detector aperture plane.    

Fourth moment $\Gamma_4({\bf r},{\bf r'})$ characterize spatiotemporal properties of the laser beam in $(x,y)$-plane in atmosphere. Functions $F$, $G$, $H$ (see Appendix \ref{appendix1}) incorporate the effect of correlation for different photon trajectories with $\{{\bf r}, {\bf q}\}$ and $\{{\bf r}', {\bf q}'\}$ on intensity correlations. As it was shown in \cite{berman2006,baskov2016} such correlations of trajectories are responsible for intensity fluctuations in the range of moderate and strong turbulence.

For the case of asymptotically large distances, $z\rightarrow\infty$, cross-correlation term vanishes because ${\Delta {r}}\rightarrow\infty$ (see (\ref{A6_0})) due to randomization of the particle
displacements from the straight lines, so functions $F$, $G$, and $H$ do not depend on ${\tilde{q}}$ and $\rho$. Therefore, the values of functions are expressed via $R_b$: $F\approx\frac{1}{2}R_b^2$, $G\approx\frac{3}{4}R_b^2$, $H\approx\frac{3}{2}R_b^2$. (Free-space terms are omitted since turbulence is assumed to give dominant contribution.) Also since cross-correlation is vanished the integration over directions for $\tilde{\bf q}$ is preserved
\begin{eqnarray}\label{gamma4_i_bigZ}
\Gamma_4^{(i)}({\bf r},{\bf r}')=C\iint d\tilde{\bf q}\left[F\big(FH-{G}^2\big)\right]^{-{1}}\times\hspace{10mm}\\\exp\left\{-\frac{\left(\tilde{\bf q}-({\bf r}-{\bf r'})\frac{G}{2F}\right)^2}{(H-{G}^2/F)}\right\}
\exp\left\{-\left(\frac{{{\bf r}}^{2}}{2F}+\frac{{{\bf r'}}^{2}}{2F}\right)\right\}\nonumber,
\end{eqnarray}
\begin{eqnarray}\label{gamma4_ii_bigZ}
\Gamma_4^{(ii)}({\bf r},{\bf r}')=\hspace{60mm}\\ C\iint d\tilde{\bf q}\left[F\big(FH-{G}^2\big)\right]^{-{1}}\exp\left\{-\frac{\tilde{\bf q}^2}{(H-{G}^2/F)}\right\}\times\nonumber\\
\exp\left\{-\left(\frac{({\bf r}+{\bf r'})^{2}}{4F}\right)\right\}\exp\left\{-i2\tilde{\bf q}({\bf r}-{\bf r'})\frac{q_0}{z}\right\}\nonumber,
\end{eqnarray}
In this case it is easy to perform integration analytically. The contribution $\Gamma_4^{(i)}({\bf r},{\bf r}')$ can be expressed via average intensity, $\langle \hat{I}({\bf r})\rangle\propto\exp\left\{-\left(\frac{{{\bf r}}^{2}}{R_b^2}\right)\right\}\big/R_b^2$ (see \cite{baskov2020})  and $\Gamma_4^{(ii)}({\bf r},{\bf r}')$ has a simple Gaussian form, so
\begin{eqnarray}\label{48my}
&&\Gamma_4({\bf r},{\bf r}')=\langle\hat{I}({\bf r}, t)\rangle\langle\hat{I}({\bf r'}, t)\rangle+\\
&&C\pi\left(\frac{1}{F}\right)^2\exp\left[{-}\frac{({\bf r}{+}{\bf r'})^2}{4F}{-}\frac{({\bf r}{-}{\bf r'})^2q_0^2(H-G^2/F)}{z^2}\right],\nonumber
\end{eqnarray}
which is exactly the result of \cite{baskov2020}, taking to account that $2F\approx R_b^2$, $q_0^2(H-G^2/F)/z^2\approx\langle{\bf q}^2\rangle_T/8$, $\langle {\bf q}^2\rangle_T=4\alpha t$, and the relation between constants.

\subsection*{Applicability of approximation}
The applicability of expressions (\ref{gamma4_i}) and (\ref{gamma4_ii}) is inherent from main approximations of the approach, i.e., collisionless Boltzmann equation and the concept of photon trajectories. As it was pointed out in \cite{berman2006} the momentum of photons should be much bigger than characteristic wave vectors of turbulence. Since upper limit of the spectrum is defined by inner scale of turbulence, $l_0$, the relation $\langle {\bf q}^2\rangle_T l_0^2$ should be large enough. On the other hand, throughout the paper we account for the effect of correlation of photon trajectories, so such concept should be justified. To consider photons as particles, whose density in the $({\bf r},{\bf q})$ domain is defined by the distribution function $\hat{f}({\bf r},{\bf q},t)$, the uncertainty of the momentum ${\bf q}$ should be small \cite{baskov2016}. The value of the uncertainty can be estimated from the definition of the distribution function (\ref{1threee}) as ${\bf k}/2$. For large distances and Tatarskii spectrum such uncertainty is estimated by the relation
\begin{equation}\label{thir69}
\langle {\bf q}^2\rangle_T R_b^2\approx 15\cdot q_0^2l_0^{-2/3}C_n^4
z^4,
\end{equation}
which also should be large enough.

\section{Fluctuations of transmitted radiation}
\label{sec:trans_fluct}

For many practical cases, e.g., development classical and quantum communication,  the fluctuation of transmittance in Earth's atmosphere is a key parameter that defines the properties of atmospheric channel \cite{vasylyev,Barrios,Vetelino}. The magnitude of fluctuations is estimated via variance
\begin{equation}\label{scinteta}
\sigma_{\eta}^2=\frac{\langle\hat{\eta}^2\rangle-\langle\hat{\eta}\rangle^2}{\langle\hat{\eta}\rangle^2},
\end{equation}
where transmittance of the optical channel is defined as
\begin{equation}
\hat{\eta}=(4C\pi^3)^{-\frac{1}2}\int\limits_{\mathcal{A}}d{\bf r} \hat{I}({\bf r},t)
\end{equation}
and accounts for the finite size of detector aperture. The normalizing condition $\langle\hat{\eta}\rangle =1$ for $\mathcal{A}$ which is much larger than the beam cross section is used.

\begin{figure}[t!]	
	\centering
	\includegraphics[width=\linewidth,keepaspectratio]{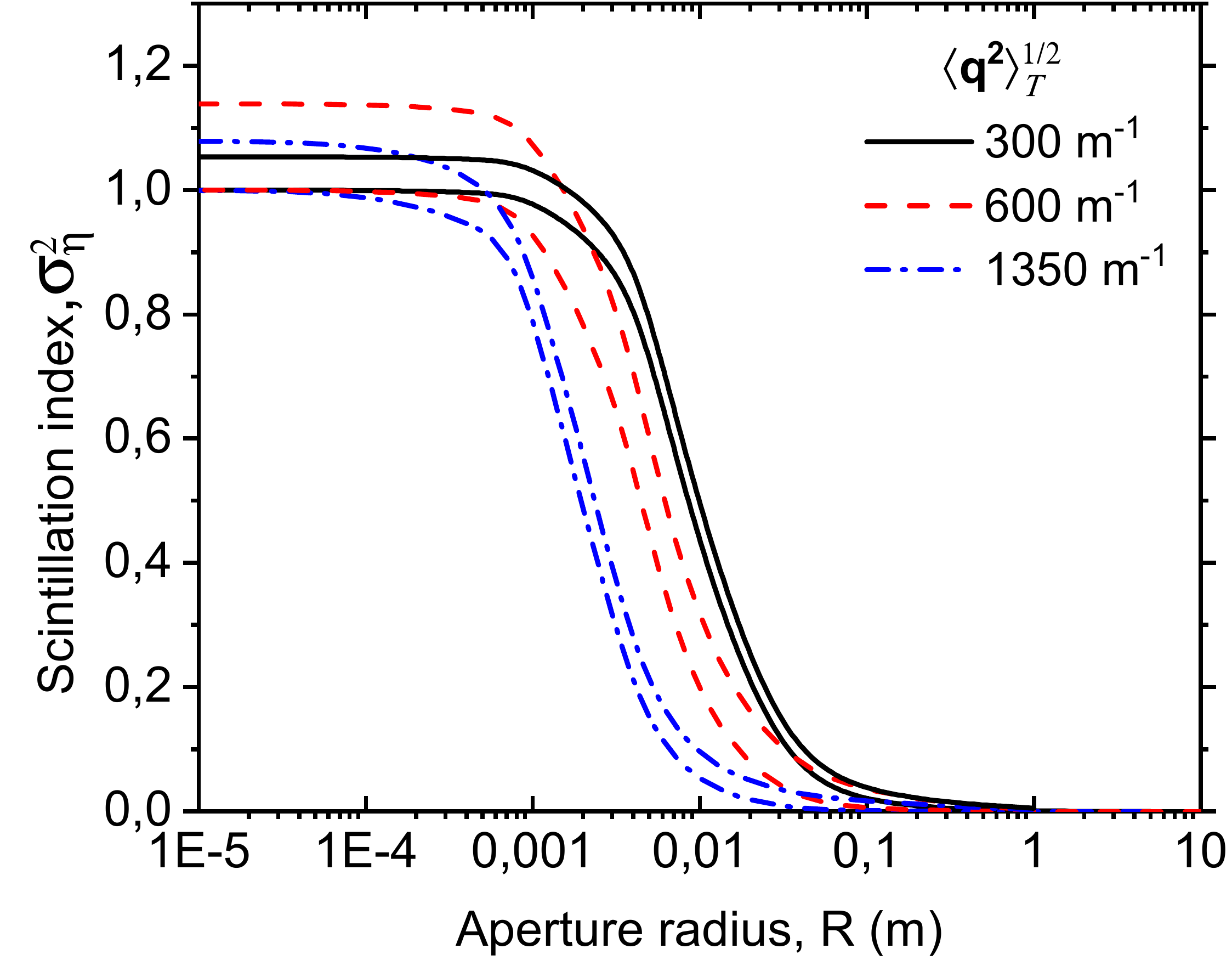}

	\caption{Effect of ``nondiagonal'' terms in $\Gamma_4$ on aperture averaged scintillations. In each pair of lines lower curve depicts the results from \cite{baskov2020}, where it is assumed that intensity fluctuations are fully saturated; upper curve represents the results for (\ref{gamma4_i}) and (\ref{gamma4_ii}). Dash-dotted lines: $z=20\,km$, $C_n^2=2.5\times10^{-14}\,m^{-2/3}$, Rytov variance $\sigma_{R}^2=62$; dashed lines: $z=17\,km$, $C_n^2=5.8\times10^{-15}\,m^{-2/3}$, $\sigma_{R}^2=60$; solid lines: $z=100\,km$, $C_n^2=2.5\times10^{-16}\,m^{-2/3}$, $\sigma_{R}^2=66$. Common parameters of the beam and channel for all curves: $r_0=0.01\,m$, $l_0/2\pi=10^{-3}\,m$, $q_0=10^{7}\,m^{-1}$.}
	\label{fig:1}
\end{figure}

Two moments  for $\eta$ are defined as \cite{fried,semenov}
\begin{eqnarray}\label{eta_avg}
\langle\hat{\eta}\rangle&=&(4C\pi^3)^{-\frac{1}2}\int\limits_{\mathcal{A}}d{\bf r} \Gamma_2({\bf r}),\\
\label{eta2_general}
\langle\hat{\eta}^2\rangle&=&(4C\pi^3)^{-1}\int\limits_{\mathcal{A}}d{\bf r} \int\limits_{\mathcal{A}}d{\bf r}^\prime \Gamma_4({\bf r},{\bf r}^\prime),
\end{eqnarray}
where $\Gamma_2({\bf r})\equiv \langle\hat{I}({\bf r}, t)\rangle$.
To obtain the fluctuations of transmitted radiation $\sigma_{\eta}^2$ we calculate numerically (five-fold integration) $\langle\hat{\eta}^2\rangle=\langle\hat{\eta}^2\rangle^{(i)}+\langle\hat{\eta}^2\rangle^{(ii)}$ for circular aperture with radius $R$ using the expressions (\ref{gamma4_i}) and (\ref{gamma4_ii}). 

\begin{figure}[t!]	
	\centering
	\includegraphics[width=\linewidth,keepaspectratio]{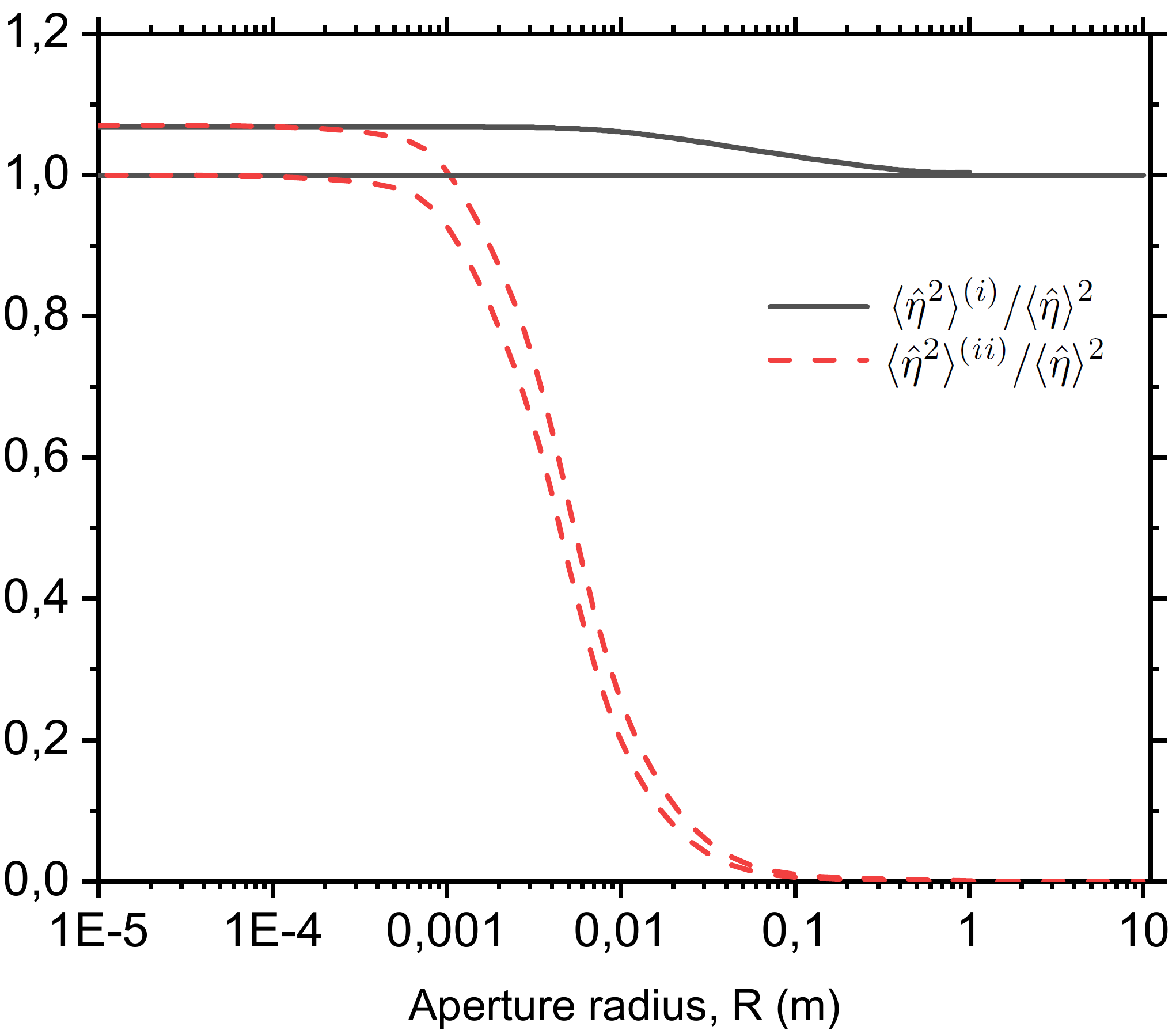}

	\caption{ Contributions of two regions $(i)$ (solid lines) and $(ii)$ (dashed lines) for partially saturated (upper pair of curves) and saturated regimes (lower pair of curves) for $\langle{\bf q}^2\rangle_T^{1/2}=600\,m^{-1}$ from Fig. \ref{fig:1}.}
	\label{fig:2}
\end{figure}

First of all, it is informative to compare approximations where partially saturated and fully saturated regimes were assumed (Fig. \ref{fig:1}) for atmospheric channels considered in \cite{baskov2020}. Such comparison allows both to estimate the accuracy of asymptotic approximation, $z\rightarrow\infty$, and to emphasize the differentiation of correlation properties for two regimes. For partially saturated radiation scintillations are slightly larger because of the additional contribution of ``nondiagonal'' terms in fourth moment which incorporate larger region of the phase space and account for residual effect of the initial statistics of radiation. Remarkably, the values of aperture averaged scintillations differ from unity for point-like aperture. That is a natural outcome for partially saturated regime where radiation still preserve some properties of initial statistics and does not fully acquire Gaussian statistics.  In contrast to approximation of fully saturated radiation there is clear dependence of the values of scintillations, $\sigma_{\eta}^2$, for small aperture sizes on Rytov parameter. In consistent with other studies, there is such size of aperture where detector could not be considered as point-like and steep reduction of the fluctuations is observed. These values strongly dependent on the transverse momentum of photons $\langle{\bf q}^2\rangle_T$. In addition, unlikely to fully saturated approximation the beam spreading $R_b^2$ also plays significant role to the behavior of aperture-averaged scintillations via both (\ref{gamma4_i}) and (\ref{gamma4_ii}). Figure \ref{fig:2} shows that present result more adequately account for the values of $\Gamma_4^{(i)}$ and corresponding correlation length. Since \cite{baskov2020} considers fully saturated regime, $\Gamma_4^{(i)}$ does not contribute to the fluctuations there. In contrast term (\ref{gamma4_i}) have sizable effect till the sizes of aperture are less than beam radius.

Also, there are two basic properties for scintillations averaging that are preserved in current approximation: for point-like detectors, ${\bf r}={\bf r'}=0$, (\ref{gamma4_i}) and (\ref{gamma4_ii}) contribute equally which repeats the result of previous works \cite{berman2006,baskov2020}; for aperture sizes reasonably larger than beam radius the fluctuations of transmittance tend to zero. 

\begin{figure}[t!]	
	\centering
	\includegraphics[width=\linewidth,keepaspectratio]{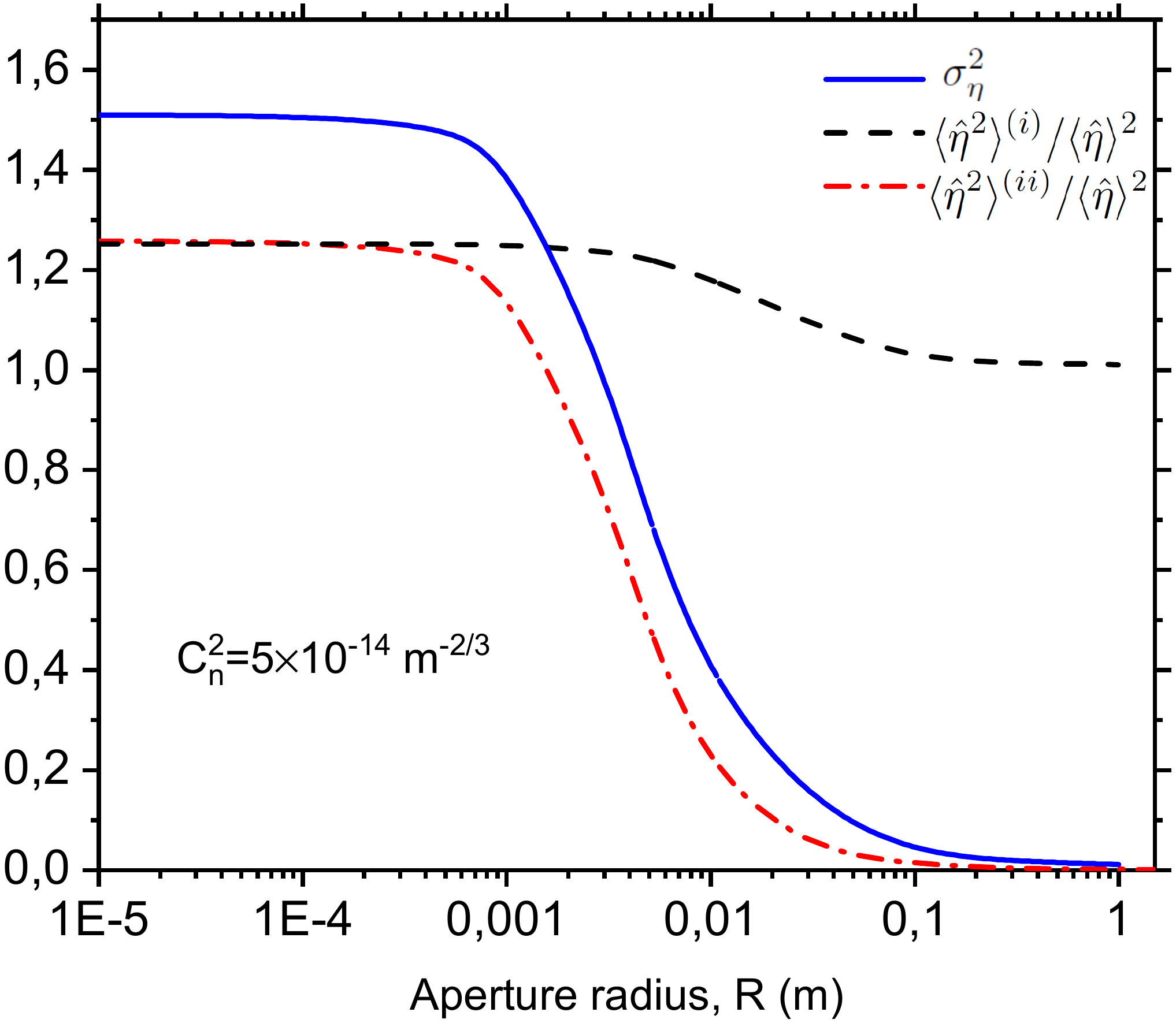}
	\includegraphics[width=\linewidth,keepaspectratio]{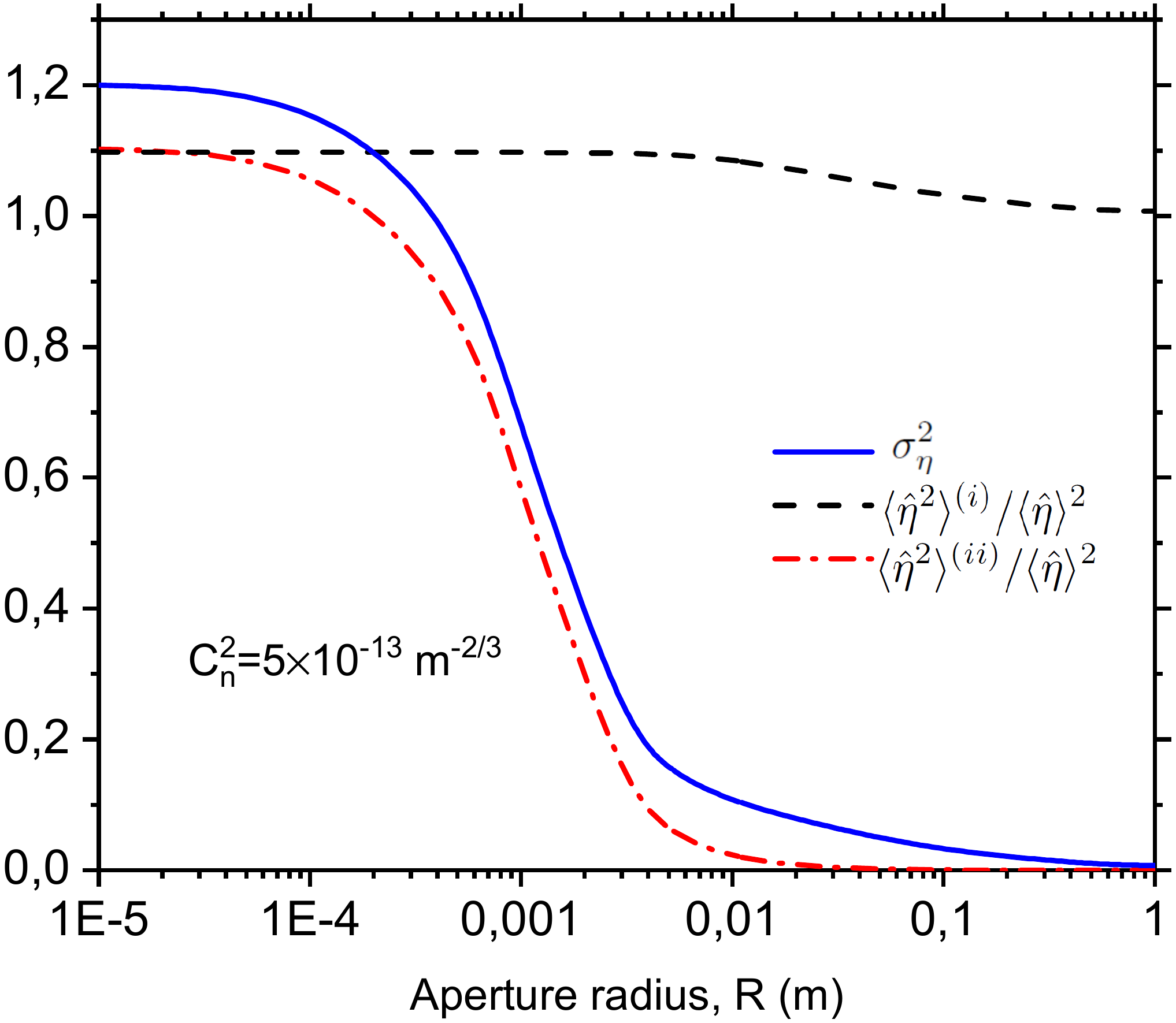}
	
	\caption{Aperture-averaged scintillation index vs. radius of detector aperture. Parameters for the both graphs: $r_0=0.01\,m$, $l_0/2\pi=10^{-3}\,m$, $q_0=1.29\times10^7\,m^{-1}$, $z=3\,km$.}
	\label{fig:3}
\end{figure}
For atmospheric channels under partially saturated regime (Fig. \ref{fig:3}) asymptotic result (\ref{48my}) significantly differs from one that considers ``nondiagonal'' terms in the expression for $\Gamma_4$. First of all, $\Gamma_4^{(i)}$ term contributes substantially to the values of transmittance fluctuations. Particularly it is responsible for long ``tails'' of the curves at large detector apertures. This effect is reminiscent to the leveling effect mentioned in \cite{Churnside1991}, where there are two characteristic scales, $\rho_0\propto\langle {\bf q}^2\rangle_T^{1/2}$ and $\rho_0z/q_0\propto R_b$, which defines correlation properties of the light radiation. Particularly one may see from Fig. \ref{fig:3} that correlation length in term $\Gamma_4^{(ii)}$ being proportional to $\langle {\bf q}^2\rangle_T^{1/2}$ is much smaller than correlation length of $\Gamma_4^{(i)}$ which comes from the values of beam radius.

\begin{figure}[t!]	
	\centering
	\includegraphics[width=\linewidth,keepaspectratio]{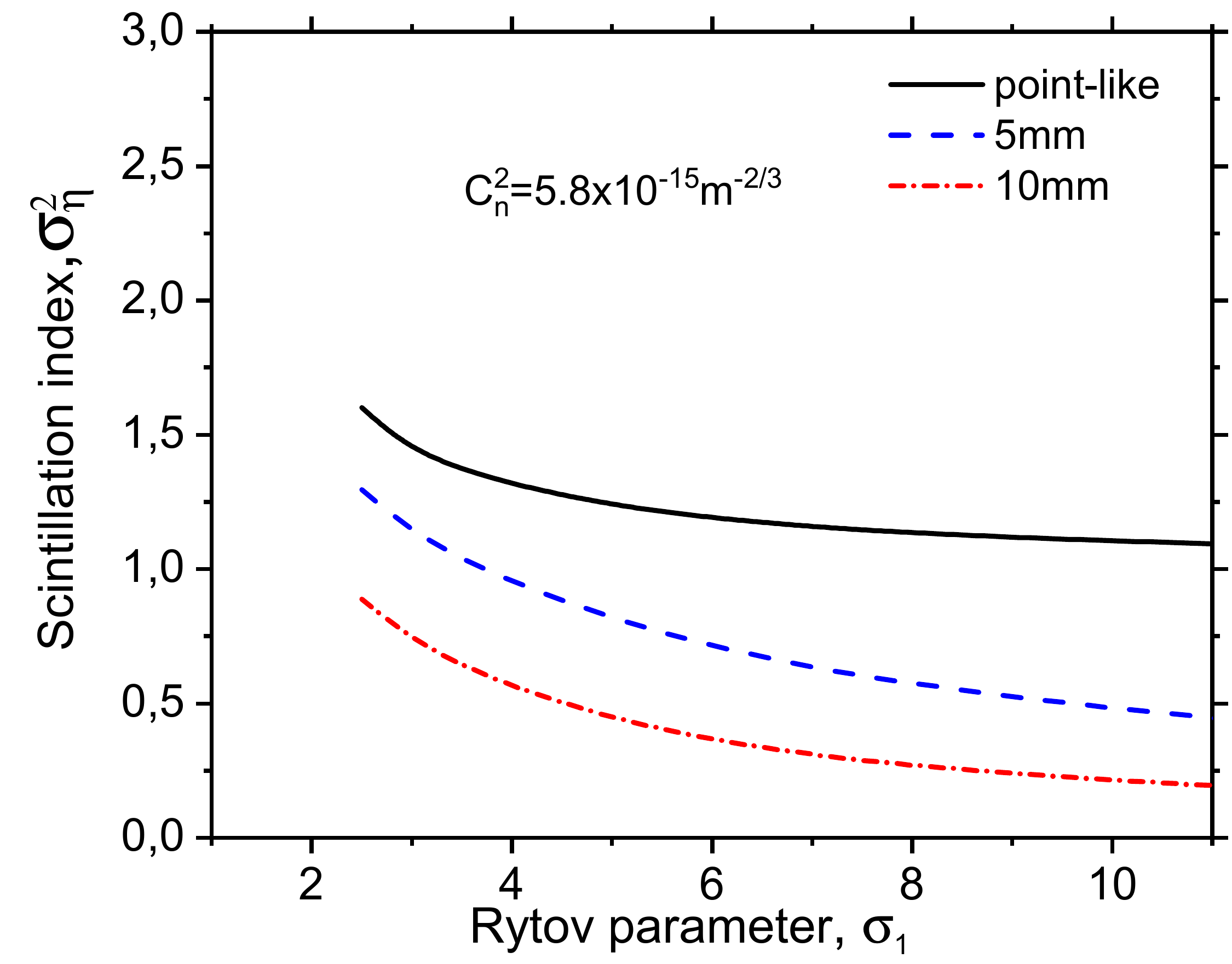}	
	
	\caption{Aperture-averaged scintillation index vs. Rytov parameter. Parameters of the channel: $r_0=0.01\,m$, $l_0/2\pi=10^{-3}\,m$, $q_0=10^7\,m^{-1}$.}
	\label{fig:4}
\end{figure}

\begin{figure}[t!]	
	\centering	
	\includegraphics[width=\linewidth,keepaspectratio]{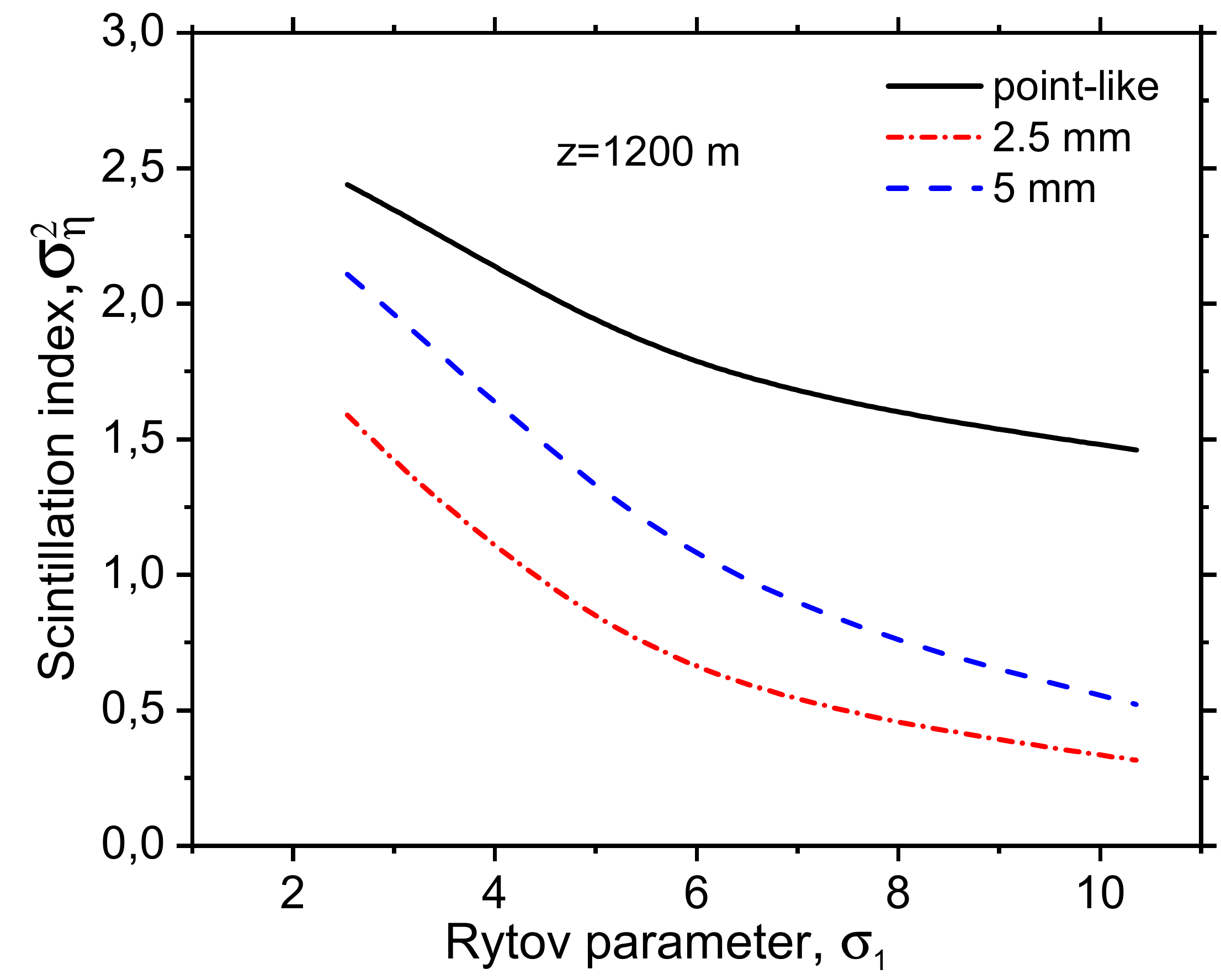}	
	\caption{Aperture-averaged scintillation index vs. Rytov parameter,  $q_0=1.29\times10^7\,m^{-1}$, other parameters are the same as in Fig. \ref{fig:3}.}
	\label{fig:5}
\end{figure}

Figures \ref{fig:4} and \ref{fig:5} depict aperture effect on scintillations in relation to values of Rytov parameter. One might see that both magnitude of fluctuations and steepness of its decrease with larger sizes of the detector aperture strongly depends on parameters of the atmospheric channel. Generally, for bigger values of Rytov parameter the effect of saturation of the fluctuations responsible for decrease of $\sigma_\eta^2$. That is, magnitude of intensity fluctuations decrease for atmospheric channels with stronger turbulence regime.

\section{Summary and Conclusions}
\label{sec:conclusion}

The fourth moment of light fields in atmosphere was derived under collisionless Boltzmann equation approximation where smooth random force represents the effect of turbulence on the laser beam. Such approximation is justified for the case of moderate and strong turbulence where fluctuations are saturated for particular degree. Randomization of the electromagnetic waves allows to distinguish two separate regions in the phase space that contribute to noise level in optical system and its correlation properties. For the case of partially saturated regime intensity correlations are not factorized to product of two-wave correlations and additional ``nondiagonal'' terms should be considered. For Gaussian beams the effect of photon trajectories correlations is enclosed in a set of functions that define distribution of photon density in the phase space. Values of such functions strongly depend on properties of the atmospheric channel. 

The fourth moment is applied for the problem of transmittance fluctuations. It is shown that current calculations describe more wide range of parameters of the atmospheric channels. There are two characteristic scales whose relation defines correlation length for detected radiation.  Found results could be useful for practical purposes.

\section*{Acknowledgments}

The author thanks D. Vasylyev, A. Semenov, and O. O. Chumak for useful discussions and comments.
The author acknowledges support from National Research Foundation of 
Ukraine through the Project 2020.02/0111.

\appendix

\begin{widetext}
\section{Calculation of $\Gamma_4$}
\label{appendix1}

In the main text we consider Gaussian beams and Tatarskii spectrum for the fluctuations of index-of-refraction. Therefore, for Gaussian beam the initial configuration of radiation defines \cite{baskov2020} 
\begin{equation}\label{A1}
\langle\phi ({\bf k},{\bf q})\rangle_{qm}=\frac {2\pi r_0^2}{VL_xL_y}\langle b^\dag b\rangle_{qm} e^{-(q_\perp^2+k_\perp^2/4)r_0^2/2},
\end{equation}
where the symbol  $\langle...\rangle_{qm}$ indicates a quantum-mechanical averaging of operators in the angle brackets.
Coefficient $C^{\prime}$ can be obtained if the total photon flux is known.  
Tatarskii spetrum is defined as
\begin{equation}\label{16my}
\psi ({\bf g})=0.033C_n^2 {\exp\left\{-(gl_0'
	)^2\right\}}{g^{-11/3}}.
\end{equation}
In the similar manner to the works  \cite{berman2006,baskov2016} we can write down the general expression for fourth moment in terms of random fluctuating force ${\bf F}_\perp$ 
\begin{eqnarray}\label{A2}
\langle I({\bf r},t)I({\bf r'},t)\rangle{=}\Bigg(\frac {2\pi r_0^2}{VL_xL_y}\Bigg)^2\langle b^\dag b^\dag bb\rangle_{qm} \sum_{\substack{ {\bf q},{\bf k},\\{\bf q^\prime},{\bf
			k^\prime}}}\bigg\langle e^{-i{\bf k}[{\bf r}-{\bf c_q}t+\frac
	c{q_0}\int_0^tdt^\prime t^\prime {\bf F}_\perp({\bf r}({\bf q}, t^\prime))]-i{\bf
		k^\prime}[{\bf r'}-{\bf c_{\bf q^\prime}} t+\frac
	c{q_0}\int_0^tdt^\prime t^\prime {\bf F}_\perp({\bf
		r}({\bf q^\prime}, t^\prime))]}\times\nonumber\\ \exp{\left(-\left(Q^2+Q'^2+\frac
	{k^2+{k^\prime} ^2}4 \right)\frac{r_0^2}2\right)}\bigg\rangle,
\end{eqnarray}
where ${\bf Q}={\bf q}-\int_0^tdt'{\bf F}_\perp({\bf r}({\bf q},t))$ and ${\bf Q'}={\bf q'}-\int_0^tdt'{\bf F}_\perp({\bf r'}({\bf q'},t))$ are the solutions of evolution equations for momenta. It is reasonable to express all turbulent part in the exponent in linear form for  ${\bf F}_\perp$. So the factor $e^{-(Q_\perp^2+{Q_\perp^\prime}^2)\frac {r_0^2}2}$ in (\ref{A2}) is expressed in the
integral form as
\begin{equation}\label{A3}
e^{-(Q_\perp^2+{Q^\prime_\perp }^2)\frac {r_0^2}2}=\int\frac{d{\bf p}d{\bf
		p^\prime}}{(2\pi r_0^2)^2}e^{i{\bf p}\cdot{\bf Q}_\perp+i{\bf p'}\cdot{\bf Q}_\perp'-(p^2+{p^\prime}^2 )/{2r_0^2}}.
\end{equation}
In this case (\ref{A2}) can be expressed as 
\begin{eqnarray}\label{A4}
\langle I({\bf r},t)I({\bf r'},t)\rangle{=}\frac {\langle b^\dag b^\dag bb\rangle_{qm}}{(VL_xL_y)^2}\int{d{\bf p}d{\bf
		p^\prime}}\sum_{\substack{ {\bf q},{\bf k},\\{\bf q^\prime},{\bf
			k^\prime}}} e^{-i{\bf k}[{\bf r}-{\bf c_q}t]-i{\bf
		k^\prime}[{\bf r'}-{\bf c_{\bf q^\prime}} t]} e^{-
	{(k^2+{k^\prime}^2)r_0^2}/{8}}e^{i{\bf p}\cdot{\bf Q}_\perp+i{\bf p'}\cdot{\bf Q}_\perp'-(p^2+{p^\prime}^2 )/{2r_0^2}}\langle M\rangle,
\end{eqnarray}
where the factor
\begin{equation}\label{A5}
M{=}\exp\left({-}i\int_0^tdt^\prime\{({\bf p}{+}{\bf k}t^\prime\frac{c}{q_0}){\bf
	F}[{\bf r}({\bf q},t^\prime )]{+}({\bf p}^\prime{+}{\bf k}^\prime
t^\prime\frac{c}{q_0}){\bf F}[{\bf r}({\bf q}^\prime,t^\prime )]\} \right)
\end{equation}
includes all fluctuating part in fourth moment. (For the sake of brevity from herein we omit $\perp$ notation for perpendicular to propagation direction.) Using similar to \cite{berman2006, baskov2016} approach, the average of $M$ is expressed as
\begin{eqnarray}\label{A6_0}
\langle M\rangle=\exp\Bigg\{ -0.033C_n^2\pi ^2q_0^2\int _0^zdx \int d{\bf g}g^{-11/3}e^{-g^2l_0'^2}
\Big[(({\bf p}+{\bf k}x /q_0)\cdot{\bf g})^2+(({\bf p}^\prime +{\bf k}^\prime x /q_0)\cdot{\bf g})^2\nonumber\\
+ 2({\bf p}+{\bf k}x /q_0)\cdot{\bf g}({\bf p}^\prime +{\bf k}^\prime x /q_0)\cdot{\bf g}
\exp\left\{i{\bf g}\cdot\boldsymbol{\Delta r}-\frac{R_b^2}{480{l_0'}^2}{\Delta r}^2(1-x/z)^{3}g^{2}\left[1+\frac{R_b^2(1-x/z)^{3}}{672{l_0'}^2}\right]\right\}\Big]\Bigg\},
\end{eqnarray}
where $g=|{\bf g}|$ and we also used Markov approximation index-of-refraction spectrum is delta-correlated
in the direction of propagation, which was rigorously justified in \cite{baskov2016}. After integration over direction of ${\bf g}$
\begin{eqnarray}\label{A6}
\langle M\rangle=\exp\Bigg\{ -0.033C_n^2\pi ^2q_0^2\int _0^zdx \int _0^\infty dgg^{-2/3}e^{-g^2l_0'^2}
\Big[({\bf p}+{\bf k}x /q_0)^2+({\bf p}^\prime +{\bf k}^\prime x /q_0)^2\hspace{20mm}\nonumber\\
 + \big(2(p_\parallel +k_\parallel x/q_0)(p_\parallel ^\prime +k_\parallel ^\prime x /q_0)
(J_0(g\Delta r)-J_2(g\Delta r))+2(p_\perp +k_\perp x/q_0)(p_\perp ^\prime +k_\perp ^\prime x/q_0)(J_0(g\Delta r)+J_2(g\Delta r))\big)\\
\times\exp\left\{-\frac{R_b^2}{480{l_0'}^2}{\Delta r}^2(1-x/z)^{3}g^{2}\left[1+\frac{R_b^2(1-x/z)^{3}}{672{l_0'}^2}\right]\right\}\Big]\Bigg\},\nonumber
\end{eqnarray}
where $J_0$ and $J_2$ are zeroth and second order Bessel functions, vector $\boldsymbol{\Delta r}=({\bf r}-{\bf r'})-({\bf q}-{\bf q}^\prime)(z-x)/q_0$, $\Delta r=|\boldsymbol{\Delta r}|$. The indices $\{_\parallel \}$ and $\{_\perp  \}$
indicate the parallel and perpendicular to $\boldsymbol{\Delta r}$ components of the corresponding 2D vectors. First two terms in square brackets represent correlation of waves with same pairs $\{{\bf r}, {\bf q}\}$ and $\{{\bf r}', {\bf q}'\}$. The  terms that include Bessel functions consider cross-correlation of photon trajectories ($\{{\bf r}, {\bf q}\}$ and $\{{\bf r}', {\bf q}'\}$ are different) which were extensively reviewed in \cite{baskov2016}. In (\ref{A6_0}) and (\ref{A6}) we omit the dependence on angle between $\boldsymbol{\Delta r}$ and ${\bf g}$ (see Eq. (40) in \cite{baskov2016}) in the last exponent as its contribution is small. After substitution of (\ref{A6}) to (\ref{A4}) and change of variables, most of integrations could be performed analytically. As a result expression for fourth moment can be written as
\begin{eqnarray}\label{A7}
\Gamma_4^{(i)}({\bf r},{\bf r}')=2\pi C\int\limits_{0}^\infty d\tilde{q}\tilde{q}\left[F_1^{\tilde{q},\rho_{\parallel}}F_2^{\tilde{q},\rho_{\parallel}}\big(F_3^{\tilde{q},\rho_{\parallel}}H_1^{\tilde{q},\rho_{\parallel}}-{G_1^{\tilde{q},\rho_{\parallel}}}^2\big)\big(F_4^{\tilde{q},\rho_{\parallel}}H_2^{\tilde{q},\rho_{\parallel}}-{G_2^{\tilde{q},\rho_{\parallel}}}^2\big)\right]^{-\frac{1}{2}}\times\hspace{30mm}\\\exp\left\{-\frac{\left(\tilde{q}-\rho_{\parallel}\frac{G_1^{\tilde{q},\rho_{\parallel}}}{2F_3^{\tilde{q},\rho_{\parallel}}}\right)^2}{(H_1^{\tilde{q},\rho_{\parallel}}-{G_1^{\tilde{q},\rho_{\parallel}}}^2/F_3^{\tilde{q},\rho_{\parallel}})}\right\}\exp\left\{-\frac{\rho_\perp^2H_2^{\tilde{q},\rho_{\parallel}}}{(H_2^{\tilde{q},\rho_{\parallel}}F_4^{\tilde{q},\rho_{\parallel}}-{G_2^{\tilde{q},\rho_{\parallel}}}^2)}\right\}
\exp\left\{-\left(\frac{\rho_{\parallel}^{\prime 2}}{F_1^{\tilde{q},\rho_{\parallel}}}+\frac{\rho_{\perp}^{\prime 2}}{F_2^{\tilde{q},\rho_{\parallel}}}+\frac{\rho_{\parallel}^{2}}{4F_3^{\tilde{q},\rho_{\parallel}}}\right)\right\}\nonumber,
\end{eqnarray}
functions $F$, $G$ and $H$ are defined as
\begin{eqnarray}
F_{1,2,3,4}=\frac{r_0^2}{4}+\frac{z^2}{q_0^2r_0^2}+\varphi_{1,2,3,4}\\
G_{1,2}=\frac{z^2}{q_0^2r_0^2}+\gamma_{1,2}\\
H_{1,2}=\frac{z^2}{q_0^2r_0^2}+\chi_{1,2}
\end{eqnarray}
where
\begin{eqnarray}
&\varphi_1{=}\alpha\int\limits_0^1d\tau\tau^2[1{+}(1{+}\beta {\Delta \tilde{r}}^2\tau^3)^{-\frac{1}{6}}[{_1F_1}(\frac{1}{6},1;\frac{-{\Delta \tilde{r}}^2}{4{l_0'}^2(1+\beta {\Delta \tilde{r}}^2\tau^2)}){+}\frac{{\Delta \tilde{r}}^2}{48{l_0'}^2(1+\beta {\Delta \tilde{r}}^2\tau^2)}{_1F_1}(\frac{7}{6},3;\frac{-{\Delta \tilde{r}}^2}{{l_0'}^2(1+\beta {\Delta \tilde{r}}^2\tau^2)})]]\label{phi1}\\
&\varphi_2{=}\alpha\int\limits_0^1d\tau\tau^2[1{+}(1{+}\beta {\Delta \tilde{r}}^2\tau^3)^{-\frac{1}{6}}[{_1F_1}(\frac{1}{6},1;\frac{-{\Delta \tilde{r}}^2}{4{l_0'}^2(1+\beta {\Delta \tilde{r}}^2\tau^2)}){-}\frac{{\Delta \tilde{r}}^2}{48{l_0'}^2(1+\beta {\Delta \tilde{r}}^2\tau^2)}{_1F_1}(\frac{7}{6},3;\frac{-{\Delta \tilde{r}}^2}{{l_0'}^2(1+\beta {\Delta \tilde{r}}^2\tau^2)})]]\\
&\varphi_3{=}\alpha\int\limits_0^1d\tau\tau^2[1{-}(1{+}\beta {\Delta \tilde{r}}^2\tau^3)^{-\frac{1}{6}}[{_1F_1}(\frac{1}{6},1;\frac{-{\Delta \tilde{r}}^2}{4{l_0'}^2(1+\beta {\Delta \tilde{r}}^2\tau^2)}){-}\frac{{\Delta \tilde{r}}^2}{48{l_0'}^2(1+\beta {\Delta \tilde{r}}^2\tau^2)}{_1F_1}(\frac{7}{6},3;\frac{-{\Delta \tilde{r}}^2}{{l_0'}^2(1+\beta {\Delta \tilde{r}}^2\tau^2)})]]\\
&\varphi_4{=}\alpha\int\limits_0^1d\tau\tau^2[1{-}(1{+}\beta {\Delta \tilde{r}}^2\tau^3)^{-\frac{1}{6}}[{_1F_1}(\frac{1}{6},1;\frac{-{\Delta \tilde{r}}^2}{4{l_0'}^2(1+\beta {\Delta \tilde{r}}^2\tau^2)}){+}\frac{{\Delta \tilde{r}}^2}{48{l_0'}^2(1+\beta {\Delta \tilde{r}}^2\tau^2)}{_1F_1}(\frac{7}{6},3;\frac{-{\Delta \tilde{r}}^2}{{l_0'}^2(1+\beta {\Delta \tilde{r}}^2\tau^2)})]]\\
&\gamma_1{=}\alpha\int\limits_0^1d\tau\tau[1{-}(1{+}\beta {\Delta \tilde{r}}^2\tau^3)^{-\frac{1}{6}}[{_1F_1}(\frac{1}{6},1;\frac{-{\Delta \tilde{r}}^2}{4{l_0'}^2(1+\beta {\Delta \tilde{r}}^2\tau^2)}){-}\frac{{\Delta \tilde{r}}^2}{48{l_0'}^2(1+\beta {\Delta \tilde{r}}^2\tau^2)}{_1F_1}(\frac{7}{6},3;\frac{-{\Delta \tilde{r}}^2}{{l_0'}^2(1+\beta {\Delta \tilde{r}}^2\tau^2)})]]\\
&\gamma_2{=}\alpha\int\limits_0^1d\tau\tau[1{-}(1{+}\beta {\Delta \tilde{r}}^2\tau^3)^{-\frac{1}{6}}[{_1F_1}(\frac{1}{6},1;\frac{-{\Delta \tilde{r}}^2}{4{l_0'}^2(1+\beta {\Delta \tilde{r}}^2\tau^2)}){+}\frac{{\Delta \tilde{r}}^2}{48{l_0'}^2(1+\beta {\Delta \tilde{r}}^2\tau^2)}{_1F_1}(\frac{7}{6},3;\frac{-{\Delta \tilde{r}}^2}{{l_0'}^2(1+\beta {\Delta \tilde{r}}^2\tau^2)})]]\\
&\chi_1{=}\alpha\int\limits_0^1d\tau[1{-}(1{+}\beta {\Delta \tilde{r}}^2\tau^3)^{-\frac{1}{6}}[{_1F_1}(\frac{1}{6},1;\frac{-{\Delta \tilde{r}}^2}{4{l_0'}^2(1+\beta {\Delta \tilde{r}}^2\tau^2)}){-}\frac{{\Delta \tilde{r}}^2}{48{l_0'}^2(1+\beta {\Delta \tilde{r}}^2\tau^2)}{_1F_1}(\frac{7}{6},3;\frac{-{\Delta \tilde{r}}^2}{{l_0'}^2(1+\beta {\Delta \tilde{r}}^2\tau^2)})]]\\
&\chi_2{=}\alpha\int\limits_0^1d\tau[1{-}(1{+}\beta {\Delta \tilde{r}}^2\tau^3)^{-\frac{1}{6}}[{_1F_1}(\frac{1}{6},1;\frac{-{\Delta \tilde{r}}^2}{4{l_0'}^2(1+\beta {\Delta \tilde{r}}^2\tau^2)}){+}\frac{{\Delta \tilde{r}}^2}{48{l_0'}^2(1+\beta {\Delta \tilde{r}}^2\tau^2)}{_1F_1}(\frac{7}{6},3;\frac{-{\Delta \tilde{r}}^2}{{l_0'}^2(1+\beta {\Delta \tilde{r}}^2\tau^2)})]]\label{chi2},
\end{eqnarray}
and parameters are defined as  $\alpha=3R_b^2$, $\beta=\frac{R_b^2}{480{l_0'}^2}\left[1+\frac{R_b^2\tau^3}{672{l_0'}^2}\right]$, ${\Delta \tilde{r}}=\rho_\parallel-2\tilde{q}\tau$ and ${}_1 F_1(a,b;z)=\sum\limits_{n=0}^{\infty}\frac{a^{(n)}z^n}{b^{(n)}n!}$ is a confluent hypergeometric function (Kummer's  function), $a^{(n)}$, $b^{(n)}$ are the Pochhammer symbols. Here notations  $\boldsymbol{\rho}={\bf r}-{\bf r}^\prime$, $\boldsymbol{\rho}^\prime=({\bf r}+{\bf r}^\prime)/2$ are used since ${\bf r}$ and ${\bf r}^\prime$ enters expression for $\Gamma_4$ only in such combinations (see more details in main text). Functions (\ref{phi1})-(\ref{chi2}) depend on $\rho_\parallel$ and $\tilde{q}$.

For the region of the phase space defined with conditions (ii) one should perform change of indices similar to one in \cite{berman2009,baskov2020}
\[
{\bf k}\rightarrow{\bf q}-{\bf q'}+\frac{{\bf k}+{\bf k'}}{2},\quad {\bf k'}\rightarrow{\bf q'}-{\bf q}+\frac{{\bf k}+{\bf k'}}{2}\label{7.1my}\]
\[{\bf q}\rightarrow\frac{1}{2}\left({\bf q}+{\bf q'}+\frac{{\bf k}-{\bf k'}}{2}\right),\quad {\bf q'}\rightarrow\frac{1}{2}\left({\bf q}+{\bf q'}-\frac{{\bf k}-{\bf k'}}{2}\right).\]
Then $\Gamma_4^{(ii)}$ is expressed as
\begin{eqnarray}\label{A8}
\Gamma_4^{(ii)}({\bf r},{\bf r'})=\sum_{{\bf q},{\bf q'}}e^{i({\bf q'}-{\bf q})\cdot({\bf r}-{\bf r'})}\langle f(\frac{{\bf r}+{\bf r'}}{2},{\bf q}) f(\frac{{\bf r}+{\bf r'}}{2},{\bf q'})\rangle,
\end{eqnarray}
which is reminiscent to the asymptotic expression for fluctuations of intensity \cite{baskov2020}. However, for smaller distances, accounting for partial saturation of fluctuations, it is not factorized to the product of first moments for PDF. Calculation of (\ref{A6}) is similar to one for term (\ref{A4}). However since PDFs in (\ref{A6}) depend on the same coordinate  $({\bf r}+{\bf r'})/{2}$ both auto-correlations and cross-correlation terms in (\ref{A6}) do not include space coordinates.  After integrations 
\begin{eqnarray}\label{A9}
\Gamma_4^{(ii)}({\bf r},{\bf r}')=2\pi C\int\limits_{0}^\infty d\tilde{q}\tilde{q}\left[F_1^{\tilde{q}}F_2^{\tilde{q}}\big(F_3^{\tilde{q}}H_1^{\tilde{q}}-{G_1^{\tilde{q}}}^{\,2}\big)\big(F_4^{\tilde{q}}H_2^{\tilde{q}}-{G_2^{\tilde{q}}}^{\,2}\big)\right]^{-\frac{1}{2}}\times\hspace{65mm}\\\exp\left\{-\frac{{\tilde{q}}^2}{H_1^{\tilde{q}}-{G_1^{\tilde{q}}}^{\,2}/F_3^{\tilde{q}}}\right\}
\exp\left\{-\left(\frac{\rho_{\parallel}^{\prime 2}}{F_1^{\tilde{q}}}+\frac{\rho_{\perp}^{\prime 2}}{F_2^{\tilde{q}}}\right)\right\}\exp\left\{-i2{\tilde{q}}\rho_{\parallel}\frac{q_0}{z}\right\}\nonumber,
\end{eqnarray}
where functions $F$,$G$ and $H$ does not depend on ${\bf r}$, ${\bf r'}$ and one should put $2\tilde{q}\tau$ instead of ${\Delta \tilde{r}}$ in corresponding expressions.
\end{widetext}

\bibliography{refs}

\begin{thebibliography}{53}%
\makeatletter
\providecommand \@ifxundefined [1]{%
 \@ifx{#1\undefined}
}%
\providecommand \@ifnum [1]{%
 \ifnum #1\expandafter \@firstoftwo
 \else \expandafter \@secondoftwo
 \fi
}%
\providecommand \@ifx [1]{%
 \ifx #1\expandafter \@firstoftwo
 \else \expandafter \@secondoftwo
 \fi
}%
\providecommand \natexlab [1]{#1}%
\providecommand \enquote  [1]{``#1''}%
\providecommand \bibnamefont  [1]{#1}%
\providecommand \bibfnamefont [1]{#1}%
\providecommand \citenamefont [1]{#1}%
\providecommand \href@noop [0]{\@secondoftwo}%
\providecommand \href [0]{\begingroup \@sanitize@url \@href}%
\providecommand \@href[1]{\@@startlink{#1}\@@href}%
\providecommand \@@href[1]{\endgroup#1\@@endlink}%
\providecommand \@sanitize@url [0]{\catcode `\\12\catcode `\$12\catcode
  `\&12\catcode `\#12\catcode `\^12\catcode `\_12\catcode `\%12\relax}%
\providecommand \@@startlink[1]{}%
\providecommand \@@endlink[0]{}%
\providecommand \url  [0]{\begingroup\@sanitize@url \@url }%
\providecommand \@url [1]{\endgroup\@href {#1}{\urlprefix }}%
\providecommand \urlprefix  [0]{URL }%
\providecommand \Eprint [0]{\href }%
\providecommand \doibase [0]{https://doi.org/}%
\providecommand \selectlanguage [0]{\@gobble}%
\providecommand \bibinfo  [0]{\@secondoftwo}%
\providecommand \bibfield  [0]{\@secondoftwo}%
\providecommand \translation [1]{[#1]}%
\providecommand \BibitemOpen [0]{}%
\providecommand \bibitemStop [0]{}%
\providecommand \bibitemNoStop [0]{.\EOS\space}%
\providecommand \EOS [0]{\spacefactor3000\relax}%
\providecommand \BibitemShut  [1]{\csname bibitem#1\endcsname}%
\let\auto@bib@innerbib\@empty
\bibitem [{\citenamefont {Capraro}\ \emph {et~al.}(2012)\citenamefont
  {Capraro}, \citenamefont {Tomaello}, \citenamefont {Dall'Arche},
  \citenamefont {Gerlin}, \citenamefont {Ursin}, \citenamefont {Vallone},\ and\
  \citenamefont {Villoresi}}]{villoresi}%
  \BibitemOpen
  \bibfield  {author} {\bibinfo {author} {\bibfnamefont {I.}~\bibnamefont
  {Capraro}}, \bibinfo {author} {\bibfnamefont {A.}~\bibnamefont {Tomaello}},
  \bibinfo {author} {\bibfnamefont {A.}~\bibnamefont {Dall'Arche}}, \bibinfo
  {author} {\bibfnamefont {F.}~\bibnamefont {Gerlin}}, \bibinfo {author}
  {\bibfnamefont {R.}~\bibnamefont {Ursin}}, \bibinfo {author} {\bibfnamefont
  {G.}~\bibnamefont {Vallone}},\ and\ \bibinfo {author} {\bibfnamefont
  {P.}~\bibnamefont {Villoresi}},\ }\href
  {https://doi.org/10.1103/PhysRevLett.109.200502} {\bibfield  {journal}
  {\bibinfo  {journal} {Phys. Rev. Lett.}\ }\textbf {\bibinfo {volume} {109}},\
  \bibinfo {pages} {200502} (\bibinfo {year} {2012})}\BibitemShut {NoStop}%
\bibitem [{\citenamefont {Usenko}\ \emph {et~al.}(2012)\citenamefont {Usenko},
  \citenamefont {Heim}, \citenamefont {Peuntinger}, \citenamefont {Wittmann},
  \citenamefont {Marquardt}, \citenamefont {Leuchs},\ and\ \citenamefont
  {Filip}}]{usenko}%
  \BibitemOpen
  \bibfield  {author} {\bibinfo {author} {\bibfnamefont {V.~C.}\ \bibnamefont
  {Usenko}}, \bibinfo {author} {\bibfnamefont {B.}~\bibnamefont {Heim}},
  \bibinfo {author} {\bibfnamefont {C.}~\bibnamefont {Peuntinger}}, \bibinfo
  {author} {\bibfnamefont {C.}~\bibnamefont {Wittmann}}, \bibinfo {author}
  {\bibfnamefont {C.}~\bibnamefont {Marquardt}}, \bibinfo {author}
  {\bibfnamefont {G.}~\bibnamefont {Leuchs}},\ and\ \bibinfo {author}
  {\bibfnamefont {R.}~\bibnamefont {Filip}},\ }\href
  {https://doi.org/10.1088/1367-2630/14/9/093048} {\bibfield  {journal}
  {\bibinfo  {journal} {New Journal of Physics}\ }\textbf {\bibinfo {volume}
  {14}},\ \bibinfo {pages} {093048} (\bibinfo {year} {2012})}\BibitemShut
  {NoStop}%
\bibitem [{\citenamefont {Hosseinidehaj}\ and\ \citenamefont
  {Malaney}(2015)}]{Hosseinidehaj}%
  \BibitemOpen
  \bibfield  {author} {\bibinfo {author} {\bibfnamefont {N.}~\bibnamefont
  {Hosseinidehaj}}\ and\ \bibinfo {author} {\bibfnamefont {R.}~\bibnamefont
  {Malaney}},\ }\href {https://doi.org/10.1103/PhysRevA.91.022304} {\bibfield
  {journal} {\bibinfo  {journal} {Phys. Rev. A}\ }\textbf {\bibinfo {volume}
  {91}},\ \bibinfo {pages} {022304} (\bibinfo {year} {2015})}\BibitemShut
  {NoStop}%
\bibitem [{\citenamefont {Aspelmeyer}\ \emph {et~al.}(2003)\citenamefont
  {Aspelmeyer}, \citenamefont {Jennewein}, \citenamefont {Pfennigbauer},
  \citenamefont {Leeb},\ and\ \citenamefont {Zeilinger}}]{Aspelmeyer}%
  \BibitemOpen
  \bibfield  {author} {\bibinfo {author} {\bibfnamefont {M.}~\bibnamefont
  {Aspelmeyer}}, \bibinfo {author} {\bibfnamefont {T.}~\bibnamefont
  {Jennewein}}, \bibinfo {author} {\bibfnamefont {M.}~\bibnamefont
  {Pfennigbauer}}, \bibinfo {author} {\bibfnamefont {W.}~\bibnamefont {Leeb}},\
  and\ \bibinfo {author} {\bibfnamefont {A.}~\bibnamefont {Zeilinger}},\ }\href
  {https://doi.org/10.1109/JSTQE.2003.820918} {\bibfield  {journal} {\bibinfo
  {journal} {IEEE Journal of Selected Topics in Quantum Electronics}\ }\textbf
  {\bibinfo {volume} {9}},\ \bibinfo {pages} {1541} (\bibinfo {year}
  {2003})}\BibitemShut {NoStop}%
\bibitem [{\citenamefont {Ma}\ \emph {et~al.}(2012)\citenamefont {Ma} \emph
  {et~al.}}]{ma}%
  \BibitemOpen
  \bibfield  {author} {\bibinfo {author} {\bibfnamefont {X.-S.}\ \bibnamefont
  {Ma}} \emph {et~al.},\ }\href {https://doi.org/10.1038/nature11472}
  {\bibfield  {journal} {\bibinfo  {journal} {Nature}\ }\textbf {\bibinfo
  {volume} {489}},\ \bibinfo {pages} {269} (\bibinfo {year}
  {2012})}\BibitemShut {NoStop}%
\bibitem [{\citenamefont {Ren}\ \emph {et~al.}(2017)\citenamefont {Ren} \emph
  {et~al.}}]{ren}%
  \BibitemOpen
  \bibfield  {author} {\bibinfo {author} {\bibfnamefont {J.-G.}\ \bibnamefont
  {Ren}} \emph {et~al.},\ }\href {https://doi.org/10.1038/nature23675}
  {\bibfield  {journal} {\bibinfo  {journal} {Nature}\ }\textbf {\bibinfo
  {volume} {549}},\ \bibinfo {pages} {70} (\bibinfo {year} {2017})}\BibitemShut
  {NoStop}%
\bibitem [{\citenamefont {Hofmann}\ \emph {et~al.}(2019)\citenamefont
  {Hofmann}, \citenamefont {Semenov}, \citenamefont {Vogel},\ and\
  \citenamefont {Bohmann}}]{Hofmann}%
  \BibitemOpen
  \bibfield  {author} {\bibinfo {author} {\bibfnamefont {K.}~\bibnamefont
  {Hofmann}}, \bibinfo {author} {\bibfnamefont {A.~A.}\ \bibnamefont
  {Semenov}}, \bibinfo {author} {\bibfnamefont {W.}~\bibnamefont {Vogel}},\
  and\ \bibinfo {author} {\bibfnamefont {M.}~\bibnamefont {Bohmann}},\ }\href
  {https://doi.org/10.1088/1402-4896/ab36e0} {\bibfield  {journal} {\bibinfo
  {journal} {Physica Scripta}\ }\textbf {\bibinfo {volume} {94}},\ \bibinfo
  {pages} {125104} (\bibinfo {year} {2019})}\BibitemShut {NoStop}%
\bibitem [{\citenamefont {{Ursin}}\ \emph {et~al.}(2007)\citenamefont {{Ursin}}
  \emph {et~al.}}]{ursin}%
  \BibitemOpen
  \bibfield  {author} {\bibinfo {author} {\bibfnamefont {R.}~\bibnamefont
  {{Ursin}}} \emph {et~al.},\ }\href {https://doi.org/10.1038/nphys629}
  {\bibfield  {journal} {\bibinfo  {journal} {Nature Physics}\ }\textbf
  {\bibinfo {volume} {3}},\ \bibinfo {pages} {481} (\bibinfo {year}
  {2007})}\BibitemShut {NoStop}%
\bibitem [{\citenamefont {Yin}\ \emph {et~al.}(2017)\citenamefont {Yin} \emph
  {et~al.}}]{yin}%
  \BibitemOpen
  \bibfield  {author} {\bibinfo {author} {\bibfnamefont {J.}~\bibnamefont
  {Yin}} \emph {et~al.},\ }\href {https://doi.org/10.1126/science.aan3211}
  {\bibfield  {journal} {\bibinfo  {journal} {Science}\ }\textbf {\bibinfo
  {volume} {356}},\ \bibinfo {pages} {1140} (\bibinfo {year}
  {2017})}\BibitemShut {NoStop}%
\bibitem [{\citenamefont {Peuntinger}\ \emph {et~al.}(2014)\citenamefont
  {Peuntinger}, \citenamefont {Heim}, \citenamefont {M\"uller}, \citenamefont
  {Gabriel}, \citenamefont {Marquardt},\ and\ \citenamefont
  {Leuchs}}]{Peuntinger}%
  \BibitemOpen
  \bibfield  {author} {\bibinfo {author} {\bibfnamefont {C.}~\bibnamefont
  {Peuntinger}}, \bibinfo {author} {\bibfnamefont {B.}~\bibnamefont {Heim}},
  \bibinfo {author} {\bibfnamefont {C.~R.}\ \bibnamefont {M\"uller}}, \bibinfo
  {author} {\bibfnamefont {C.}~\bibnamefont {Gabriel}}, \bibinfo {author}
  {\bibfnamefont {C.}~\bibnamefont {Marquardt}},\ and\ \bibinfo {author}
  {\bibfnamefont {G.}~\bibnamefont {Leuchs}},\ }\href
  {https://doi.org/10.1103/PhysRevLett.113.060502} {\bibfield  {journal}
  {\bibinfo  {journal} {Phys. Rev. Lett.}\ }\textbf {\bibinfo {volume} {113}},\
  \bibinfo {pages} {060502} (\bibinfo {year} {2014})}\BibitemShut {NoStop}%
\bibitem [{\citenamefont {Vasylyev}\ \emph {et~al.}(2016)\citenamefont
  {Vasylyev}, \citenamefont {Semenov},\ and\ \citenamefont
  {Vogel}}]{vasylev2016}%
  \BibitemOpen
  \bibfield  {author} {\bibinfo {author} {\bibfnamefont {D.}~\bibnamefont
  {Vasylyev}}, \bibinfo {author} {\bibfnamefont {A.~A.}\ \bibnamefont
  {Semenov}},\ and\ \bibinfo {author} {\bibfnamefont {W.}~\bibnamefont
  {Vogel}},\ }\href {https://doi.org/10.1103/PhysRevLett.117.090501} {\bibfield
   {journal} {\bibinfo  {journal} {Phys. Rev. Lett.}\ }\textbf {\bibinfo
  {volume} {117}},\ \bibinfo {pages} {090501} (\bibinfo {year}
  {2016})}\BibitemShut {NoStop}%
\bibitem [{\citenamefont {Meyers}(2015)}]{Meyers}%
  \BibitemOpen
  \bibfield  {author} {\bibinfo {author} {\bibfnamefont {R.~E.}\ \bibnamefont
  {Meyers}},\ }\bibinfo {title} {Free-space and atmospheric quantum
  communications},\ in\ \href {https://doi.org/10.1007/978-1-4939-0918-6_10}
  {\emph {\bibinfo {booktitle} {Advanced Free Space Optics (FSO): A Systems
  Approach}}}\ (\bibinfo  {publisher} {Springer New York},\ \bibinfo {address}
  {New York, NY},\ \bibinfo {year} {2015})\ pp.\ \bibinfo {pages}
  {343--387}\BibitemShut {NoStop}%
\bibitem [{\citenamefont {Cao}\ \emph {et~al.}(2020)\citenamefont {Cao} \emph
  {et~al.}}]{cao2020}%
  \BibitemOpen
  \bibfield  {author} {\bibinfo {author} {\bibfnamefont {Y.}~\bibnamefont
  {Cao}} \emph {et~al.},\ }\href
  {https://doi.org/10.1103/PhysRevLett.125.260503} {\bibfield  {journal}
  {\bibinfo  {journal} {Phys. Rev. Lett.}\ }\textbf {\bibinfo {volume} {125}},\
  \bibinfo {pages} {260503} (\bibinfo {year} {2020})}\BibitemShut {NoStop}%
\bibitem [{\citenamefont {Kish}\ \emph {et~al.}(2020)\citenamefont {Kish},
  \citenamefont {Villasenor}, \citenamefont {Malaney}, \citenamefont {Mudge},\
  and\ \citenamefont {Grant}}]{kish}%
  \BibitemOpen
  \bibfield  {author} {\bibinfo {author} {\bibfnamefont {S.~P.}\ \bibnamefont
  {Kish}}, \bibinfo {author} {\bibfnamefont {E.}~\bibnamefont {Villasenor}},
  \bibinfo {author} {\bibfnamefont {R.}~\bibnamefont {Malaney}}, \bibinfo
  {author} {\bibfnamefont {K.~A.}\ \bibnamefont {Mudge}},\ and\ \bibinfo
  {author} {\bibfnamefont {K.~J.}\ \bibnamefont {Grant}},\ }\href
  {https://doi.org/https://doi.org/10.1002/que2.50} {\bibfield  {journal}
  {\bibinfo  {journal} {Quantum Engineering}\ }\textbf {\bibinfo {volume}
  {2}},\ \bibinfo {pages} {e50} (\bibinfo {year} {2020})}\BibitemShut {NoStop}%
\bibitem [{\citenamefont {Fante}(1975)}]{Fante1}%
  \BibitemOpen
  \bibfield  {author} {\bibinfo {author} {\bibfnamefont {R.}~\bibnamefont
  {Fante}},\ }\href {https://doi.org/10.1109/PROC.1975.10035} {\bibfield
  {journal} {\bibinfo  {journal} {Proceedings of the IEEE}\ }\textbf {\bibinfo
  {volume} {63}},\ \bibinfo {pages} {1669} (\bibinfo {year}
  {1975})}\BibitemShut {NoStop}%
\bibitem [{\citenamefont {Fante}(1980)}]{Fante2}%
  \BibitemOpen
  \bibfield  {author} {\bibinfo {author} {\bibfnamefont {R.}~\bibnamefont
  {Fante}},\ }\href {https://doi.org/10.1109/PROC.1980.11882} {\bibfield
  {journal} {\bibinfo  {journal} {Proceedings of the IEEE}\ }\textbf {\bibinfo
  {volume} {68}},\ \bibinfo {pages} {1424} (\bibinfo {year}
  {1980})}\BibitemShut {NoStop}%
\bibitem [{\citenamefont {Tatarskii}(1971)}]{Tatarskii1}%
  \BibitemOpen
  \bibfield  {author} {\bibinfo {author} {\bibfnamefont {V.~I.}\ \bibnamefont
  {Tatarskii}},\ }\href@noop {} {\emph {\bibinfo {title} {The effects of the
  turbulent atmosphere on wave propagation}}}\ (\bibinfo  {publisher} {Israel
  Program for Scientific Translations ; Reproduced by National Technical
  Information Service, U.S. Dept. of Commerce},\ \bibinfo {address} {Jerusalem;
  Springfield, VA.},\ \bibinfo {year} {1971})\BibitemShut {NoStop}%
\bibitem [{\citenamefont {Andrews}\ and\ \citenamefont
  {Phillips}(2005)}]{Andrews}%
  \BibitemOpen
  \bibfield  {author} {\bibinfo {author} {\bibfnamefont {L.}~\bibnamefont
  {Andrews}}\ and\ \bibinfo {author} {\bibfnamefont {R.}~\bibnamefont
  {Phillips}},\ }\href {https://books.google.com.ua/books?id=4NXHYg70qqIC}
  {\emph {\bibinfo {title} {Laser Beam Propagation Through Random Media}}},\
  Online access with subscription: SPIE Digital Library\ (\bibinfo  {publisher}
  {Society of Photo Optical},\ \bibinfo {year} {2005})\BibitemShut {NoStop}%
\bibitem [{\citenamefont {DeWolf}(1968)}]{DeWolf}%
  \BibitemOpen
  \bibfield  {author} {\bibinfo {author} {\bibfnamefont {D.~A.}\ \bibnamefont
  {DeWolf}},\ }\href {https://doi.org/10.1364/JOSA.58.000461} {\bibfield
  {journal} {\bibinfo  {journal} {J. Opt. Soc. Am.}\ }\textbf {\bibinfo
  {volume} {58}},\ \bibinfo {pages} {461} (\bibinfo {year} {1968})}\BibitemShut
  {NoStop}%
\bibitem [{\citenamefont {Gracheva}\ and\ \citenamefont
  {Gurvich}(1965)}]{gracheva}%
  \BibitemOpen
  \bibfield  {author} {\bibinfo {author} {\bibfnamefont {M.~E.}\ \bibnamefont
  {Gracheva}}\ and\ \bibinfo {author} {\bibfnamefont {A.~S.}\ \bibnamefont
  {Gurvich}},\ }\href {https://doi.org/10.1007/BF01038327} {\bibfield
  {journal} {\bibinfo  {journal} {Soviet Radiophysics}\ }\textbf {\bibinfo
  {volume} {8}},\ \bibinfo {pages} {511} (\bibinfo {year} {1965})}\BibitemShut
  {NoStop}%
\bibitem [{\citenamefont {Andrews}\ \emph {et~al.}(1997)\citenamefont
  {Andrews}, \citenamefont {Phillips},\ and\ \citenamefont
  {Weeks}}]{andrews1997}%
  \BibitemOpen
  \bibfield  {author} {\bibinfo {author} {\bibfnamefont {L.~C.}\ \bibnamefont
  {Andrews}}, \bibinfo {author} {\bibfnamefont {R.~L.}\ \bibnamefont
  {Phillips}},\ and\ \bibinfo {author} {\bibfnamefont {A.~R.}\ \bibnamefont
  {Weeks}},\ }\href {https://doi.org/10.1364/JOSAA.14.001938} {\bibfield
  {journal} {\bibinfo  {journal} {J. Opt. Soc. Am. A}\ }\textbf {\bibinfo
  {volume} {14}},\ \bibinfo {pages} {1938} (\bibinfo {year}
  {1997})}\BibitemShut {NoStop}%
\bibitem [{\citenamefont {Dashen}(1979)}]{Dashen}%
  \BibitemOpen
  \bibfield  {author} {\bibinfo {author} {\bibfnamefont {R.}~\bibnamefont
  {Dashen}},\ }\href {https://doi.org/10.1063/1.524138} {\bibfield  {journal}
  {\bibinfo  {journal} {Journal of Mathematical Physics}\ }\textbf {\bibinfo
  {volume} {20}},\ \bibinfo {pages} {894} (\bibinfo {year} {1979})}\BibitemShut
  {NoStop}%
\bibitem [{\citenamefont {Banakh}\ and\ \citenamefont
  {Mironov}(1979)}]{banakh79}%
  \BibitemOpen
  \bibfield  {author} {\bibinfo {author} {\bibfnamefont {V.~A.}\ \bibnamefont
  {Banakh}}\ and\ \bibinfo {author} {\bibfnamefont {V.~L.}\ \bibnamefont
  {Mironov}},\ }\href {https://doi.org/10.1364/OL.4.000259} {\bibfield
  {journal} {\bibinfo  {journal} {Opt. Lett.}\ }\textbf {\bibinfo {volume}
  {4}},\ \bibinfo {pages} {259} (\bibinfo {year} {1979})}\BibitemShut {NoStop}%
\bibitem [{\citenamefont {Berman}\ and\ \citenamefont
  {Chumak}(2006)}]{berman2006}%
  \BibitemOpen
  \bibfield  {author} {\bibinfo {author} {\bibfnamefont {G.~P.}\ \bibnamefont
  {Berman}}\ and\ \bibinfo {author} {\bibfnamefont {A.~A.}\ \bibnamefont
  {Chumak}},\ }\href {https://doi.org/10.1103/PhysRevA.74.013805} {\bibfield
  {journal} {\bibinfo  {journal} {Phys. Rev. A}\ }\textbf {\bibinfo {volume}
  {74}},\ \bibinfo {pages} {013805} (\bibinfo {year} {2006})}\BibitemShut
  {NoStop}%
\bibitem [{\citenamefont {Berman}\ and\ \citenamefont
  {Chumak}(2009)}]{berman2009}%
  \BibitemOpen
  \bibfield  {author} {\bibinfo {author} {\bibfnamefont {G.~P.}\ \bibnamefont
  {Berman}}\ and\ \bibinfo {author} {\bibfnamefont {A.~A.}\ \bibnamefont
  {Chumak}},\ }\href {https://doi.org/10.1103/PhysRevA.79.063848} {\bibfield
  {journal} {\bibinfo  {journal} {Phys. Rev. A}\ }\textbf {\bibinfo {volume}
  {79}},\ \bibinfo {pages} {063848} (\bibinfo {year} {2009})}\BibitemShut
  {NoStop}%
\bibitem [{\citenamefont {Berman}\ \emph {et~al.}(2007)\citenamefont {Berman},
  \citenamefont {Chumak},\ and\ \citenamefont {Gorshkov}}]{Gorshkov}%
  \BibitemOpen
  \bibfield  {author} {\bibinfo {author} {\bibfnamefont {G.~P.}\ \bibnamefont
  {Berman}}, \bibinfo {author} {\bibfnamefont {A.~A.}\ \bibnamefont {Chumak}},\
  and\ \bibinfo {author} {\bibfnamefont {V.~N.}\ \bibnamefont {Gorshkov}},\
  }\href {https://doi.org/10.1103/PhysRevE.76.056606} {\bibfield  {journal}
  {\bibinfo  {journal} {Phys. Rev. E}\ }\textbf {\bibinfo {volume} {76}},\
  \bibinfo {pages} {056606} (\bibinfo {year} {2007})}\BibitemShut {NoStop}%
\bibitem [{\citenamefont {Chumak}\ and\ \citenamefont
  {Stolyarov}(2013)}]{stolyarov2013}%
  \BibitemOpen
  \bibfield  {author} {\bibinfo {author} {\bibfnamefont {O.~O.}\ \bibnamefont
  {Chumak}}\ and\ \bibinfo {author} {\bibfnamefont {E.~V.}\ \bibnamefont
  {Stolyarov}},\ }\href {https://doi.org/10.1103/PhysRevA.88.013855} {\bibfield
   {journal} {\bibinfo  {journal} {Phys. Rev. A}\ }\textbf {\bibinfo {volume}
  {88}},\ \bibinfo {pages} {013855} (\bibinfo {year} {2013})}\BibitemShut
  {NoStop}%
\bibitem [{\citenamefont {Chumak}\ and\ \citenamefont
  {Stolyarov}(2014)}]{stolyarov2014}%
  \BibitemOpen
  \bibfield  {author} {\bibinfo {author} {\bibfnamefont {O.~O.}\ \bibnamefont
  {Chumak}}\ and\ \bibinfo {author} {\bibfnamefont {E.~V.}\ \bibnamefont
  {Stolyarov}},\ }\href {https://doi.org/10.1103/PhysRevA.90.063832} {\bibfield
   {journal} {\bibinfo  {journal} {Phys. Rev. A}\ }\textbf {\bibinfo {volume}
  {90}},\ \bibinfo {pages} {063832} (\bibinfo {year} {2014})}\BibitemShut
  {NoStop}%
\bibitem [{\citenamefont {Baskov}\ and\ \citenamefont
  {Chumak}(2018)}]{baskov2018}%
  \BibitemOpen
  \bibfield  {author} {\bibinfo {author} {\bibfnamefont {R.~A.}\ \bibnamefont
  {Baskov}}\ and\ \bibinfo {author} {\bibfnamefont {O.~O.}\ \bibnamefont
  {Chumak}},\ }\href {https://doi.org/10.1103/PhysRevA.97.043817} {\bibfield
  {journal} {\bibinfo  {journal} {Phys. Rev. A}\ }\textbf {\bibinfo {volume}
  {97}},\ \bibinfo {pages} {043817} (\bibinfo {year} {2018})}\BibitemShut
  {NoStop}%
\bibitem [{\citenamefont {Chumak}\ and\ \citenamefont
  {Baskov}(2016)}]{baskov2016}%
  \BibitemOpen
  \bibfield  {author} {\bibinfo {author} {\bibfnamefont {O.~O.}\ \bibnamefont
  {Chumak}}\ and\ \bibinfo {author} {\bibfnamefont {R.~A.}\ \bibnamefont
  {Baskov}},\ }\href {https://doi.org/10.1103/PhysRevA.93.033821} {\bibfield
  {journal} {\bibinfo  {journal} {Phys. Rev. A}\ }\textbf {\bibinfo {volume}
  {93}},\ \bibinfo {pages} {033821} (\bibinfo {year} {2016})}\BibitemShut
  {NoStop}%
\bibitem [{\citenamefont {Baskov}\ and\ \citenamefont
  {Chumak}(2020)}]{baskov2020}%
  \BibitemOpen
  \bibfield  {author} {\bibinfo {author} {\bibfnamefont {R.~A.}\ \bibnamefont
  {Baskov}}\ and\ \bibinfo {author} {\bibfnamefont {O.~O.}\ \bibnamefont
  {Chumak}},\ }\href {https://doi.org/10.1088/2040-8986/abb2f1} {\bibfield
  {journal} {\bibinfo  {journal} {Journal of Optics}\ }\textbf {\bibinfo
  {volume} {22}},\ \bibinfo {pages} {105603} (\bibinfo {year}
  {2020})}\BibitemShut {NoStop}%
\bibitem [{\citenamefont {Vellekoop}\ \emph {et~al.}(2010)\citenamefont
  {Vellekoop}, \citenamefont {Lagendijk},\ and\ \citenamefont
  {Mosk}}]{Vellekoop2010}%
  \BibitemOpen
  \bibfield  {author} {\bibinfo {author} {\bibfnamefont {I.~M.}\ \bibnamefont
  {Vellekoop}}, \bibinfo {author} {\bibfnamefont {A.}~\bibnamefont
  {Lagendijk}},\ and\ \bibinfo {author} {\bibfnamefont {A.~P.}\ \bibnamefont
  {Mosk}},\ }\href {https://doi.org/10.1038/nphoton.2010.3} {\bibfield
  {journal} {\bibinfo  {journal} {Nature Photonics}\ }\textbf {\bibinfo
  {volume} {4}},\ \bibinfo {pages} {320} (\bibinfo {year} {2010})}\BibitemShut
  {NoStop}%
\bibitem [{\citenamefont {Newman}\ and\ \citenamefont {Webb}(2014)}]{Newman}%
  \BibitemOpen
  \bibfield  {author} {\bibinfo {author} {\bibfnamefont {J.~A.}\ \bibnamefont
  {Newman}}\ and\ \bibinfo {author} {\bibfnamefont {K.~J.}\ \bibnamefont
  {Webb}},\ }\href {https://doi.org/10.1103/PhysRevLett.113.263903} {\bibfield
  {journal} {\bibinfo  {journal} {Phys. Rev. Lett.}\ }\textbf {\bibinfo
  {volume} {113}},\ \bibinfo {pages} {263903} (\bibinfo {year}
  {2014})}\BibitemShut {NoStop}%
\bibitem [{\citenamefont {Popoff}\ \emph {et~al.}(2014)\citenamefont {Popoff},
  \citenamefont {Goetschy}, \citenamefont {Liew}, \citenamefont {Stone},\ and\
  \citenamefont {Cao}}]{Popoff}%
  \BibitemOpen
  \bibfield  {author} {\bibinfo {author} {\bibfnamefont {S.~M.}\ \bibnamefont
  {Popoff}}, \bibinfo {author} {\bibfnamefont {A.}~\bibnamefont {Goetschy}},
  \bibinfo {author} {\bibfnamefont {S.~F.}\ \bibnamefont {Liew}}, \bibinfo
  {author} {\bibfnamefont {A.~D.}\ \bibnamefont {Stone}},\ and\ \bibinfo
  {author} {\bibfnamefont {H.}~\bibnamefont {Cao}},\ }\href
  {https://doi.org/10.1103/PhysRevLett.112.133903} {\bibfield  {journal}
  {\bibinfo  {journal} {Phys. Rev. Lett.}\ }\textbf {\bibinfo {volume} {112}},\
  \bibinfo {pages} {133903} (\bibinfo {year} {2014})}\BibitemShut {NoStop}%
\bibitem [{\citenamefont {Vellekoop}\ and\ \citenamefont
  {Mosk}(2007)}]{Vellekoop2007}%
  \BibitemOpen
  \bibfield  {author} {\bibinfo {author} {\bibfnamefont {I.~M.}\ \bibnamefont
  {Vellekoop}}\ and\ \bibinfo {author} {\bibfnamefont {A.~P.}\ \bibnamefont
  {Mosk}},\ }\href {https://doi.org/10.1364/OL.32.002309} {\bibfield  {journal}
  {\bibinfo  {journal} {Opt. Lett.}\ }\textbf {\bibinfo {volume} {32}},\
  \bibinfo {pages} {2309} (\bibinfo {year} {2007})}\BibitemShut {NoStop}%
\bibitem [{\citenamefont {{Katz}}\ \emph {et~al.}(2012)\citenamefont {{Katz}},
  \citenamefont {{Small}},\ and\ \citenamefont {{Silberberg}}}]{Katz}%
  \BibitemOpen
  \bibfield  {author} {\bibinfo {author} {\bibfnamefont {O.}~\bibnamefont
  {{Katz}}}, \bibinfo {author} {\bibfnamefont {E.}~\bibnamefont {{Small}}},\
  and\ \bibinfo {author} {\bibfnamefont {Y.}~\bibnamefont {{Silberberg}}},\
  }\href {https://doi.org/10.1038/nphoton.2012.150} {\bibfield  {journal}
  {\bibinfo  {journal} {Nature Photonics}\ }\textbf {\bibinfo {volume} {6}},\
  \bibinfo {pages} {549} (\bibinfo {year} {2012})}\BibitemShut {NoStop}%
\bibitem [{\citenamefont {Hardy}\ and\ \citenamefont {Shapiro}(2013)}]{Hardy}%
  \BibitemOpen
  \bibfield  {author} {\bibinfo {author} {\bibfnamefont {N.~D.}\ \bibnamefont
  {Hardy}}\ and\ \bibinfo {author} {\bibfnamefont {J.~H.}\ \bibnamefont
  {Shapiro}},\ }\href {https://doi.org/10.1103/PhysRevA.87.023820} {\bibfield
  {journal} {\bibinfo  {journal} {Phys. Rev. A}\ }\textbf {\bibinfo {volume}
  {87}},\ \bibinfo {pages} {023820} (\bibinfo {year} {2013})}\BibitemShut
  {NoStop}%
\bibitem [{\citenamefont {Zhang}\ \emph {et~al.}(2010)\citenamefont {Zhang},
  \citenamefont {Gong}, \citenamefont {Shen},\ and\ \citenamefont
  {Han}}]{Zhang_imag}%
  \BibitemOpen
  \bibfield  {author} {\bibinfo {author} {\bibfnamefont {P.}~\bibnamefont
  {Zhang}}, \bibinfo {author} {\bibfnamefont {W.}~\bibnamefont {Gong}},
  \bibinfo {author} {\bibfnamefont {X.}~\bibnamefont {Shen}},\ and\ \bibinfo
  {author} {\bibfnamefont {S.}~\bibnamefont {Han}},\ }\href
  {https://doi.org/10.1103/PhysRevA.82.033817} {\bibfield  {journal} {\bibinfo
  {journal} {Phys. Rev. A}\ }\textbf {\bibinfo {volume} {82}},\ \bibinfo
  {pages} {033817} (\bibinfo {year} {2010})}\BibitemShut {NoStop}%
\bibitem [{\citenamefont {Wang}\ \emph {et~al.}(2010)\citenamefont {Wang},
  \citenamefont {Cai},\ and\ \citenamefont {Korotkova}}]{wang}%
  \BibitemOpen
  \bibfield  {author} {\bibinfo {author} {\bibfnamefont {F.}~\bibnamefont
  {Wang}}, \bibinfo {author} {\bibfnamefont {Y.}~\bibnamefont {Cai}},\ and\
  \bibinfo {author} {\bibfnamefont {O.}~\bibnamefont {Korotkova}},\ }in\ \href
  {https://doi.org/10.1117/12.841278} {\emph {\bibinfo {booktitle} {Atmospheric
  and Oceanic Propagation of Electromagnetic Waves IV}}},\ Vol.\ \bibinfo
  {volume} {7588},\ \bibinfo {editor} {edited by\ \bibinfo {editor}
  {\bibfnamefont {O.}~\bibnamefont {Korotkova}}},\ \bibinfo {organization}
  {International Society for Optics and Photonics}\ (\bibinfo  {publisher}
  {SPIE},\ \bibinfo {year} {2010})\ pp.\ \bibinfo {pages} {137 --
  144}\BibitemShut {NoStop}%
\bibitem [{\citenamefont {Shi}\ \emph {et~al.}(2013)\citenamefont {Shi},
  \citenamefont {Fan}, \citenamefont {Zhang}, \citenamefont {Shen},
  \citenamefont {Zhang}, \citenamefont {Qiao},\ and\ \citenamefont
  {Wang}}]{Shi}%
  \BibitemOpen
  \bibfield  {author} {\bibinfo {author} {\bibfnamefont {D.}~\bibnamefont
  {Shi}}, \bibinfo {author} {\bibfnamefont {C.}~\bibnamefont {Fan}}, \bibinfo
  {author} {\bibfnamefont {P.}~\bibnamefont {Zhang}}, \bibinfo {author}
  {\bibfnamefont {H.}~\bibnamefont {Shen}}, \bibinfo {author} {\bibfnamefont
  {J.}~\bibnamefont {Zhang}}, \bibinfo {author} {\bibfnamefont
  {C.}~\bibnamefont {Qiao}},\ and\ \bibinfo {author} {\bibfnamefont
  {Y.}~\bibnamefont {Wang}},\ }\href {https://doi.org/10.1364/OE.21.002050}
  {\bibfield  {journal} {\bibinfo  {journal} {Opt. Express}\ }\textbf {\bibinfo
  {volume} {21}},\ \bibinfo {pages} {2050} (\bibinfo {year}
  {2013})}\BibitemShut {NoStop}%
\bibitem [{\citenamefont {Semenov}\ and\ \citenamefont
  {Vogel}(2009)}]{Semenov2009}%
  \BibitemOpen
  \bibfield  {author} {\bibinfo {author} {\bibfnamefont {A.~A.}\ \bibnamefont
  {Semenov}}\ and\ \bibinfo {author} {\bibfnamefont {W.}~\bibnamefont
  {Vogel}},\ }\href {https://doi.org/10.1103/PhysRevA.80.021802} {\bibfield
  {journal} {\bibinfo  {journal} {Phys. Rev. A}\ }\textbf {\bibinfo {volume}
  {80}},\ \bibinfo {pages} {021802} (\bibinfo {year} {2009})}\BibitemShut
  {NoStop}%
\bibitem [{\citenamefont {Xue}\ \emph {et~al.}(2020)\citenamefont {Xue},
  \citenamefont {Shi}, \citenamefont {Chen}, \citenamefont {Yin}, \citenamefont
  {Fan-Yuan}, \citenamefont {Fu}, \citenamefont {Lu},\ and\ \citenamefont
  {Wei}}]{Xue2020}%
  \BibitemOpen
  \bibfield  {author} {\bibinfo {author} {\bibfnamefont {Y.}~\bibnamefont
  {Xue}}, \bibinfo {author} {\bibfnamefont {L.}~\bibnamefont {Shi}}, \bibinfo
  {author} {\bibfnamefont {W.}~\bibnamefont {Chen}}, \bibinfo {author}
  {\bibfnamefont {Z.}~\bibnamefont {Yin}}, \bibinfo {author} {\bibfnamefont
  {G.-j.}\ \bibnamefont {Fan-Yuan}}, \bibinfo {author} {\bibfnamefont
  {H.}~\bibnamefont {Fu}}, \bibinfo {author} {\bibfnamefont {Q.}~\bibnamefont
  {Lu}},\ and\ \bibinfo {author} {\bibfnamefont {J.}~\bibnamefont {Wei}},\
  }\href {https://doi.org/10.1103/PhysRevA.102.062602} {\bibfield  {journal}
  {\bibinfo  {journal} {Phys. Rev. A}\ }\textbf {\bibinfo {volume} {102}},\
  \bibinfo {pages} {062602} (\bibinfo {year} {2020})}\BibitemShut {NoStop}%
\bibitem [{\citenamefont {Tarasenko}\ \emph {et~al.}(1992)\citenamefont
  {Tarasenko}, \citenamefont {Tomchuk},\ and\ \citenamefont {Chumak}}]{Tom}%
  \BibitemOpen
  \bibfield  {author} {\bibinfo {author} {\bibfnamefont {A.~A.}\ \bibnamefont
  {Tarasenko}}, \bibinfo {author} {\bibfnamefont {P.~M.}\ \bibnamefont
  {Tomchuk}},\ and\ \bibinfo {author} {\bibfnamefont {A.~A.}\ \bibnamefont
  {Chumak}},\ }\href@noop {} {\emph {\bibinfo {title} {Fluctuations in the bulk
  and on the surface of solids}}}\ (\bibinfo  {publisher} {Naukova Dumka},\
  \bibinfo {address} {Kyiv},\ \bibinfo {year} {1992})\ \bibinfo {note} {(in
  Russian)}\BibitemShut {NoStop}%
\bibitem [{\citenamefont {Rarenko}\ \emph {et~al.}(1992)\citenamefont
  {Rarenko}, \citenamefont {Tarasenko}, ,\ and\ \citenamefont
  {Chumak}}]{ujp1992}%
  \BibitemOpen
  \bibfield  {author} {\bibinfo {author} {\bibfnamefont {A.~I.}\ \bibnamefont
  {Rarenko}}, \bibinfo {author} {\bibfnamefont {A.~A.}\ \bibnamefont
  {Tarasenko}}, ,\ and\ \bibinfo {author} {\bibfnamefont {A.~A.}\ \bibnamefont
  {Chumak}},\ }\href@noop {} {\bibfield  {journal} {\bibinfo  {journal} {Ukr.
  J. Phys.}\ }\textbf {\bibinfo {volume} {37}},\ \bibinfo {pages} {1577}
  (\bibinfo {year} {1992})}\BibitemShut {NoStop}%
\bibitem [{\citenamefont {Chumak}\ and\ \citenamefont
  {Sushkova}(2012)}]{ujp2012}%
  \BibitemOpen
  \bibfield  {author} {\bibinfo {author} {\bibfnamefont {O.}~\bibnamefont
  {Chumak}}\ and\ \bibinfo {author} {\bibfnamefont {N.}~\bibnamefont
  {Sushkova}},\ }\href@noop {} {\bibfield  {journal} {\bibinfo  {journal} {Ukr.
  J. Phys.}\ }\textbf {\bibinfo {volume} {57}},\ \bibinfo {pages} {30}
  (\bibinfo {year} {2012})}\BibitemShut {NoStop}%
\bibitem [{\citenamefont {Strohbehn}\ and\ \citenamefont
  {Clifford}(1967)}]{stroh}%
  \BibitemOpen
  \bibfield  {author} {\bibinfo {author} {\bibfnamefont {J.}~\bibnamefont
  {Strohbehn}}\ and\ \bibinfo {author} {\bibfnamefont {S.}~\bibnamefont
  {Clifford}},\ }\href {https://doi.org/10.1109/TAP.1967.1138937} {\bibfield
  {journal} {\bibinfo  {journal} {IEEE Transactions on Antennas and
  Propagation}\ }\textbf {\bibinfo {volume} {15}},\ \bibinfo {pages} {416}
  (\bibinfo {year} {1967})}\BibitemShut {NoStop}%
\bibitem [{\citenamefont {Landau}\ and\ \citenamefont
  {Lifshitz}(1960)}]{landau}%
  \BibitemOpen
  \bibfield  {author} {\bibinfo {author} {\bibfnamefont {L.~D.}\ \bibnamefont
  {Landau}}\ and\ \bibinfo {author} {\bibfnamefont {E.~M.}\ \bibnamefont
  {Lifshitz}},\ }\href@noop {} {\emph {\bibinfo {title} {Electrodynamics of
  continuous media vol. 8.}}}\ (\bibinfo  {publisher} {Pergamon Press},\
  \bibinfo {address} {London},\ \bibinfo {year} {1960})\BibitemShut {NoStop}%
\bibitem [{\citenamefont {Vasylyev}\ \emph {et~al.}(2019)\citenamefont
  {Vasylyev}, \citenamefont {Vogel},\ and\ \citenamefont {Moll}}]{vasylyev}%
  \BibitemOpen
  \bibfield  {author} {\bibinfo {author} {\bibfnamefont {D.}~\bibnamefont
  {Vasylyev}}, \bibinfo {author} {\bibfnamefont {W.}~\bibnamefont {Vogel}},\
  and\ \bibinfo {author} {\bibfnamefont {F.}~\bibnamefont {Moll}},\ }\href
  {https://doi.org/10.1103/PhysRevA.99.053830} {\bibfield  {journal} {\bibinfo
  {journal} {Phys. Rev. A}\ }\textbf {\bibinfo {volume} {99}},\ \bibinfo
  {pages} {053830} (\bibinfo {year} {2019})}\BibitemShut {NoStop}%
\bibitem [{\citenamefont {Barrios}\ and\ \citenamefont {Dios}(2012)}]{Barrios}%
  \BibitemOpen
  \bibfield  {author} {\bibinfo {author} {\bibfnamefont {R.}~\bibnamefont
  {Barrios}}\ and\ \bibinfo {author} {\bibfnamefont {F.}~\bibnamefont {Dios}},\
  }\href {https://doi.org/10.1364/OE.20.013055} {\bibfield  {journal} {\bibinfo
   {journal} {Opt. Express}\ }\textbf {\bibinfo {volume} {20}},\ \bibinfo
  {pages} {13055} (\bibinfo {year} {2012})}\BibitemShut {NoStop}%
\bibitem [{\citenamefont {Vetelino}\ \emph {et~al.}(2007)\citenamefont
  {Vetelino}, \citenamefont {Young}, \citenamefont {Andrews},\ and\
  \citenamefont {Recolons}}]{Vetelino}%
  \BibitemOpen
  \bibfield  {author} {\bibinfo {author} {\bibfnamefont {F.~S.}\ \bibnamefont
  {Vetelino}}, \bibinfo {author} {\bibfnamefont {C.}~\bibnamefont {Young}},
  \bibinfo {author} {\bibfnamefont {L.}~\bibnamefont {Andrews}},\ and\ \bibinfo
  {author} {\bibfnamefont {J.}~\bibnamefont {Recolons}},\ }\href
  {https://doi.org/10.1364/AO.46.002099} {\bibfield  {journal} {\bibinfo
  {journal} {Appl. Opt.}\ }\textbf {\bibinfo {volume} {46}},\ \bibinfo {pages}
  {2099} (\bibinfo {year} {2007})}\BibitemShut {NoStop}%
\bibitem [{\citenamefont {Fried}(1967)}]{fried}%
  \BibitemOpen
  \bibfield  {author} {\bibinfo {author} {\bibfnamefont {D.~L.}\ \bibnamefont
  {Fried}},\ }\href {https://doi.org/10.1364/JOSA.57.000169} {\bibfield
  {journal} {\bibinfo  {journal} {J. Opt. Soc. Am.}\ }\textbf {\bibinfo
  {volume} {57}},\ \bibinfo {pages} {169} (\bibinfo {year} {1967})}\BibitemShut
  {NoStop}%
\bibitem [{\citenamefont {Vasylyev}\ \emph {et~al.}(2018)\citenamefont
  {Vasylyev}, \citenamefont {Vogel},\ and\ \citenamefont {Semenov}}]{semenov}%
  \BibitemOpen
  \bibfield  {author} {\bibinfo {author} {\bibfnamefont {D.}~\bibnamefont
  {Vasylyev}}, \bibinfo {author} {\bibfnamefont {W.}~\bibnamefont {Vogel}},\
  and\ \bibinfo {author} {\bibfnamefont {A.~A.}\ \bibnamefont {Semenov}},\
  }\href {https://doi.org/10.1103/PhysRevA.97.063852} {\bibfield  {journal}
  {\bibinfo  {journal} {Phys. Rev. A}\ }\textbf {\bibinfo {volume} {97}},\
  \bibinfo {pages} {063852} (\bibinfo {year} {2018})}\BibitemShut {NoStop}%
\bibitem [{\citenamefont {Churnside}(1991)}]{Churnside1991}%
  \BibitemOpen
  \bibfield  {author} {\bibinfo {author} {\bibfnamefont {J.~H.}\ \bibnamefont
  {Churnside}},\ }\href {https://doi.org/10.1364/AO.30.001982} {\bibfield
  {journal} {\bibinfo  {journal} {Appl. Opt.}\ }\textbf {\bibinfo {volume}
  {30}},\ \bibinfo {pages} {1982} (\bibinfo {year} {1991})}\BibitemShut
  {NoStop}%
\end{thebibliography}%

\end{document}